\documentclass[useAMS,usenatbib]{mn2e}
\usepackage{graphicx}
\usepackage{amsmath,amssymb}
\newcommand{\del}{\partial}

\newcommand{\THE}{{\boldsymbol{\theta}}}
\newcommand{\ALP}{{\boldsymbol{\alpha}}}

\newcommand{\x}{{\boldsymbol{x}}}
\newcommand{\y}{{\boldsymbol{y}}}

\newcommand{\f}{\frac}

\newcommand{\BF}{\begin{figure}\begin{center}}
\newcommand{\EF}{\end{center}\end{figure}}
\newcommand{\BE}{\begin{equation}}
\newcommand{\EE}{\end{equation}}
\newcommand{\BEA}{\begin{eqnarray}}
\newcommand{\EEA}{\end{eqnarray}}

\newcommand{\ms}{M_{\odot}}
\begin{document}
\title{Weak lensing by line-of-sight halos 
as the origin of flux-ratio anomalies in quadruply lensed QSOs  }
\author[Kaiki Taro Inoue and Ryuichi Takahashi]
{Kaiki Taro Inoue$^{1,2}$\thanks{E-mail:kinoue@phys.kindai.ac.jp}
\thanks{E-mail:Kaiki.Inoue@astro.ox.ac.uk}
and Ryuichi Takahashi$^3$
\\
$^{1}$Department of Science and Engineering, 
Kinki University, Higashi-Osaka, 577-8502, Japan 
\\
$^{2}$Department of Physics, University of Oxford, Denys Wilkinson
Building, Keble Road, OX1 3RH, UK
\\ 
$^{3}$Faculty of Science and Technology, Hirosaki University, 3
Bunkyo-cho, 
Hirosaki, Aomori 036-8561, Japan }


\date{\today}

\pagerange{\pageref{firstpage}--\pageref{lastpage}} \pubyear{0000}

\maketitle

\label{firstpage}
\begin{abstract}
We explore the weak lensing effect by line-of-sight halos and sub-halos 
with a mass of $M \lesssim 10^7\,\ms$ in Quasi-Stellar
 Object(QSO)-galaxy 
strong lens systems with quadruple
images in a concordant $\Lambda$ cold dark matter universe.  
Using a polynomially fitted non-linear power spectrum
$P(k)$ obtained from $N$-body simulations that can resolve halos with a
 mass of $M \sim 10^5 \ms$, or structures with a comoving wavenumber 
of $k\sim 3\times 10^2\,h \textrm{Mpc}^{-1}$, we find that the ratio of 
magnification perturbation due to intervening halos
to that of a primary lens is typically $\sim 10$ per cent  
and the predicted values agree well with the estimated values 
for 6 observed QSO-galaxy lens systems with quadruple images in the
 mid-infrared band without considering the effects of substructures inside a primary lens. 
We also find that the estimated amplitudes of convergence perturbation
for the 6 lenses increase with the source redshift as predicted by 
theoretical models. Using an extrapolated matter power spectrum, 
we demonstrate that small halos or sub-halos in the line-of-sight with a
 mass of $M=10^3-10^7 \ms$, or structures with 
a comoving wavenumber of 
$k=3\times 10^2-10^4\, h \textrm{Mpc}^{-1}$ can significantly affect the magnification
 ratios of the lensed images. Flux ratio anomalies in 
QSO-galaxy strong lens systems offer us a unique probe 
into clustering property of minihalos with a mass of
$M < 10^6 \, \ms$. 
\end{abstract}

\section{Introduction}
Gravitational lensing is one of the most powerful tools for directly
probing the structure and the distribution of dark matter.
The remarkable agreement between the predicted and the observed
weak lensing effects by large-scale structures or clusters
provides independent and consistent estimates of clustering property
of dark matter on cosmic scales $\gtrsim 10\, h^{-1} \textrm{Mpc}$. 
However, we do not fully understand the clustering property 
on scales below $\sim 1\,h^{-1} \textrm{Mpc}$, which correspond to 
individual galaxy halos. Although the cold dark matter (CDM)
model predicts a large population of mini-halos ($\lesssim 10^7\, \ms$),
the observed number of dwarf galaxies in our galaxy seems too
low in comparison with the predicted value.  The discrepancy may be
alleviated by some baryonic process, such as suppression of star
formation by background UV radiation in the reionization epoch (e.g.,
\citet{bullock2000}, \citet{busha2010}), or tidal disruption due to a galactic disk \citep{donghia2010}. 
Alternatively, the suppression of the number count might be associated with
super-weakly interacting massive particles (super-WIMPs) 
or warm dark matter which has a larger free-streaming length
than CDM \citep{hisano2006}.  
In order to probe the clustering property of dark matter at mass scales
of $\lesssim 1\,h^{-1} \textrm{Mpc}$, strong QSO-galaxy lensing systems 
with quadruple images 
have been used in literature \citep{metcalf2001,chiba2002}.
In fact, the flux ratios in some quadruply lensed
QSOs disagree with the prediction of best-fit lens models with a potential 
whose fluctuation scale is larger than the separation between the
lensed images. Such a discrepancy called the ``anomalous flux ratio''
has been considered as an imprint of substructure inside a lensing galaxy
\citep{mao1998,metcalf2001,metcalf2004,chiba2005,sugai2007,mckean2007,
more2009,minezaki2009,macleod2009}.

However, recent studies based on high resolution simulations 
suggested that the predicted substructure population is too low 
to explain the observed anomalous flux ratios
\citep{maccio2006,amara2006,
xu2009,xu2010,chen2009,chen2011}. 
More detailed modeling of gravitational potential of
the lens on scales comparable to or larger than the distance between the
lensed images might also improve the fit \citep{wong2011}. However, the origin of the 
anomalous flux ratios in some quadruple image systems such as B1422+231
and MG0414+0534 has been veiled in mystery \citep{chiba2005, minezaki2009}. 

In addition to substructures in lensing galaxy, 
any intergalactic halos along the entire line-of-sight from the source
to the observer can perturb the lensing potential. Therefore, they may
change the flux ratios of the lensed images. \citet{chen2003} have found
that the contribution from intergalactic halos modeled as singular
isothermal spheres would be $\lesssim10\,\%$ of that from substructures within the lensing halo. 
\citet{metcalf2005a} performed ray-tracing simulations for intergalactic
halos with a mass of $10^6\, \ms \le M \le 10^9\, \ms$. Assuming
that the halos have Navarro, Frenk \& White (NFW) \citep{navarro1997} profiles and the
number density is given by the Press-Schechter mass function \citep{press1974}, he found
that four radio lensed QSOs that shows
a strong cusp-caustic violation are consistent with the predicted values
without any substructures in the lensing galaxy. 
Assuming that halo profiles are modeled as singular isothermal spheres 
and the number density is given by the Sheth-Tormen mass function 
\citep{sheth2002},
\citet{miranda2007} obtained a similar conclusion for 
three radio and two optical/IR lensed QSOs. 
Using a $N$-body simulation that can resolve halos with a mass of $> 10^8\,
h^{-1} \ms$, and halos with a mass ($10^6\,\ms \le M \le \, 10^8\, \ms$) whose
number density obeys the Sheth-Tormen mass function, \citet{xu2012} 
obtained a result that violation of
the cusp-caustic relation caused by line-of-sight halos are comparable
to (even larger than) those caused by intrinsic substructures though
it depends sensitively on the halo profile.

In order to estimate the magnification perturbation due to intervening
halos more precisely, it is important to take into account various effects that have
been overlooked in literature. Firstly, if the shifts in relative positions 
of images and lens due to line-of-sight halos are too large, 
fitting a model with a smooth potential 
to the observed data becomes difficult since such a change is a
consequence of a local effect. 
Moreover, even if the individual perturbing halo is not so
massive, clustering halos could produce larger image shifts. 
Therefore, we need to incorporate the effects of 
clustering as well as the shifts of position of images and lens. 
Accuracy in observed positions of lensed images and 
lens would give an upper limit on the mass scale of perturbing halos. 
Secondly, in some lens systems, violation of the cusp-caustic relation might be caused
by relatively massive faint satellite galaxies in the neighborhood of the lensing galaxy
\citep{mckean2007,shin2008,macleod2009}. 
Therefore, application of the cusp-caustic relation 
to generic lensed QSO systems may not be appropriate. Instead, we need
to use other statistics to fit the model. 
Thirdly, the effects of massive line-of-sight halos
should be subtracted off since they can contribute 
to low-order components in magnification tensor such as 
a constant convergence and an external shear in the lens model.  
Otherwise, we would estimate anomalies in the flux ratios systematically
large because of double counting.

In this paper, we explore the weak lensing effect due to line-of-sight
halos in QSO-galaxy lensing systems taking these three effects into
account and study how it will affect
the flux ratios of lensed QSOs with quadruple images.
To take into account of halo clustering, we use $N$-body simulations to calculate the
non-linear power spectrum of matter fluctuations down to 
mass scales of $\sim 10^5\, h^{-1 } \ms$. For simplicity, however, we do not 
put baryons in our $N$-body simulations.
Then we estimate the magnification perturbation using the obtained non-linear power spectrum and 
study wheather observed lensed QSO systems with quadruple images 
are consistent with 
our model prediction. In section 2, we describe magnification
perturbation due to line-of-sight halos. In section 3, we derive 
analytic formulae for the 
power spectrum of convergence due to line-of-sight halos 
constrained from perturbations in image shifts. In section 4, 
we describe our $N$-body simulations for obtaining the 
non-linear power spectrum. In section 5, image shifts and 
magnification perturbation are investigated using 
a semi-analytic method developed in section 3. 
In Section 6, we describe 6 samples of QSO-galaxy lensing
systems with quadruple images 
observed in the mid infrared (MIR) band.
In section 7, we present our
results on the flux ratio anomalies using these lens sysetems. 
In section 8, we conclude and 
discuss some relevant issues. In what follows, we assume a 
cosmology with a matter density $\Omega_m=0.272$, a baryon density 
$\Omega_b=0.046$, a cosmological constant $\Omega_\Lambda=0.728$,
the Hubble constant $H_0=70, \textrm{km}/\textrm{s}/\textrm{Mpc}$,
the spectrum index $n_s=0.97$, and the root-mean-square (rms) 
amplitude of matter fluctuations at $8 h^{-1}\, \textrm{Mpc}$, 
$\sigma_8=0.81$, which are obtained from the observed 
CMB (WMAP 7yr result, \citep{jarosik2011}), the baryon 
acoustic oscillations (Percival et~al. 2010), and $H_0$ \citep{riess2009}.    
 
\section{Perturbation of Magnification }
\begin{figure}
\vspace*{0.3cm}
\includegraphics[width=90mm]{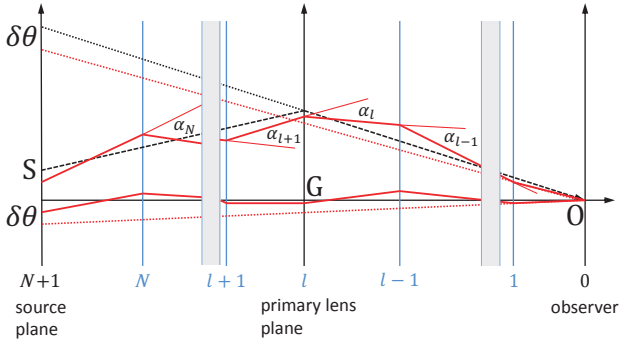}
\caption{Schematic diagram of ray tracing through multiple lens
 planes. Unperturbed light ray (dashed lines) starts 
from a source S and is deflected by the 
primary lens galaxy G at the primary lens (the $l$th) plane
and ends at O, a position of an observer. The 
light ray is perturbed by intergalactic halos in the line-of-sight (full
 curves).
$\hat{\alpha}_n$ is the deflection angle at the $n$th lens 
plane for $n=1,2,\cdots, N$. 
$\epsilon $ is the shift of image position 
or that of the center of primary lens. }
\label{f1}
\vspace*{0.2cm}
\end{figure}

Suppose a QSO at redshift $z_S$ is lensed by a primary 
lensing galaxy at $z_L$ to produce multiple images $X_i$
and less massive intergalactic halos (secondary lenses)  
perturb the QSO-galaxy lens system.
In what follows, we assume that the size of a light source is sufficiently small
in comparison with the Einstein radius of the primary lens and
those of perturbers in the line-of-sight.
Choosing coordinates centered at a primary lens, 
given the angular position of a point on the source $\THE_y$, the
angular position of the source $\THE_x$ lensed by the primary
lens and the intergalactic halos is 
approximately given by the lens equation defined at 
multiple lens planes $n=1,2,\cdots, N$ (see Fig. \ref{f1}),
\BE
D_S \THE_y=D_S \THE_x-\sum_{n=1}^{N} D_{n,S} \hat{\ALP}_i(\x_n),
\label{eq:lens1}
\EE   
where $D_S$ and $D_{n,S}$ are the angular diameter distances 
between an observer and the source, the $n$th lens plane and the source,
respectively, and $\hat{\ALP}_n$ and $\x_n$ are the deflection angle caused by a perturber
and the two-dimensional position vector in the proper coordinates 
at the $n$th lens plane, respectively. $\x_n$'s satisfy 
\BE
\x_n= \left\{
\begin{array}{ll}
D_{01}\THE_x \,, &  n=1\\
D_{0i} \THE_x-\sum_{m=1}^{n-1} D_{m,n} \hat{\ALP}_{m}(\x_m),\, &  1<n\le N, \\
\end{array}
\right.
\label{eq:lens2}
\EE
where $D_{01}$ and $D_{n,m}$ are the angular diameter distances between an observer and
the first lens plane, and between the $n$th lens and the $m$th lens
planes, respectively. We assume that the primary lens
is placed at the $l$th plane. In general, it is difficult to solve
equations (\ref{eq:lens1}) and (\ref{eq:lens2}) since the position $\x_n$ depends on any other
positions $\x_m,\, m \ne n$. However, if the
spatial derivatives of deflection angles are sufficiently small that
\BE
\frac{|\it{\Delta} \hat{\ALP}_n|} {|\hat{\ALP}_n|}= \biggl | \frac{\del \hat{\ALP}_n} {\del \x_n}
\cdot \hat{\ALP}_n \biggr | \frac{D_{n,n+1}}{|\hat{\ALP}_n| }
\ll 1,
\EE
is satisfied, then the light ray approximately follows an unperturbed
geodesic
\BE
\x_n \approx \left\{
\begin{array}{ll}
D_{0n}\THE_x \,, &  n \le l\\
D_{0n} \THE_x-D_{l,n} \hat{\ALP}_{l}(D_L \THE_x),\, & n > l, \\
\end{array}
\right.
\label{eq:lens2app}
\EE
and 
each deflection angle $\hat{\ALP}_n$ depends only 
on the position vector $\x_n$ that is independent
of the other position vectors. This greatly
simplifies the lens equations (\ref{eq:lens1}) and (\ref{eq:lens2}) 
since each $\hat{\ALP}_n$ becomes independent each other. In what
follows, we assume that equation (3) holds in our lensing systems.
Then the inverse of
the magnification tensor is approximately given by
\BE
M^{-1}=\frac{\del \THE_y}{\del \THE_x} \approx \boldsymbol{1}-
\frac{\del \ALP_l(\x_l)}{\del \THE_x}-
\sum_{n\ne l}\frac{\del \ALP_n(\x_n)}{\del \THE_x},
\EE     
where $\ALP_n=D_{n,N+1}D_S^{-1} \hat{\ALP_n}$.

In terms of convergence $\kappa_i$
and shear $\gamma_{i1}$ and $\gamma_{i2}$ due to the primary lens 
at the position of a point-like lensed image $X_i$ where $i$ denotes
the index number of lensed images, 
the contribution from the primary lens can be written as  
\BE
\Gamma_i=\frac{\del \ALP_l(\x_l)}{\del \THE_x} =\left[\begin{array}{cc}
    \kappa_i+\gamma_{i1} &   \gamma_{i2}   \\
      \gamma_{i2}   & \kappa_i-\gamma_{i1} \\
    \end{array}\right]. 
\EE
In a similar manner, in terms of perturbations of convergence $\delta \kappa$
and shear $\delta \gamma_1$, $\delta \gamma_2$, the 
contribution from clustering dark matter in the line-of-sight
can be written as 
\BE
\delta \Gamma=
\sum_{i\ne l}\frac{\del \ALP_i(\x_i)}{\del \THE_x}= \left[\begin{array}{cc}
    \delta \kappa+\delta \gamma_1 &   \delta \gamma_2   \\
     \delta \gamma_2   & \delta \kappa-\delta \gamma_1 \\
    \end{array}\right]. 
\EE
The approximated lens equation at each image position is then
\BE
\THE_y=(\boldsymbol{1}-\Gamma_i-\delta \Gamma)\THE_x.
\EE 
In the following, we assume that perturbations of flux of images 
due to shifts of positions are sufficiently 
smaller than those due to distortion of the images. 
Then the perturbed magnification matrix is given by
\BEA
(\mu_i+\delta \mu_i)^{-1}&=&(1-\kappa_i-\gamma_{i1}-\delta \kappa-\delta \gamma_1)
\nonumber
\\
&\times&(1-\kappa_i+\gamma_{i1}-\delta \kappa+\delta \gamma_1) -(\gamma_{i2}+\delta \gamma_2)^2,
\nonumber
\\
\EEA
where $\mu_i^{-1}
=(1-\kappa_i)^2-\gamma_{i1}^2-\gamma_{i2}^2$. A magnification contrast for image $X_i$
is defined by $\delta_i^{\mu}\equiv \delta \mu_i/\mu_i$. 

Up to linear order in $\mu_i \delta \kappa, \mu_i \delta \gamma_1,$ and
$\mu_i \delta \gamma_2$,
the magnification contrast is approximated as 
\BE
\delta_i^{\mu}\approx \frac{2(1-\kappa_i)\delta \kappa + 2\gamma_{i1}
\delta \gamma_1+2 \gamma_{i2}\delta\gamma_2 }{(1-\kappa_i)^2-(\gamma_{i1}^2+\gamma_{i2}^2)},
\label{eq:delta}
\EE
which can be written as
\BE
\delta_i^{\mu}\approx \frac{2(1-\kappa_i)\delta \kappa + 2\gamma_i
\delta \gamma_1}{(1-\kappa_i)^2-\gamma_i^2},
\label{eq:delta2}
\EE
if magnification matrix for the primary lens is diagonalized (i.e., $\gamma_{i2}=0$).

We expect that line-of-sight structures
that significantly perturb the fluxes of images 
are relatively massive halos, which add $\delta \kappa>0 $ to 
the background convergence. Assuming matter 
fluctuations that are homogeneous and isotropic, the mean of the shear
perturbation is
vanishing but the root-mean-square value is expected to be \citep{bartelmann2001}
\BE
\sqrt{\langle (\delta \gamma_1)^2 \rangle} =\sqrt{\langle (\delta
\gamma_2)^2 \rangle} = \sqrt{\langle (\delta \kappa)^2 \rangle}/\sqrt{2}.
\EE
Therefore, in what follows, we assume that the shear perturbations satisfy 
\BE 
-\delta \kappa/\sqrt{2}<\delta \gamma_j < \delta \kappa/\sqrt{2},~~~ j=1,2.
\EE 

The sign of magnification contrast depends on the curvature of the 
arrival time surface where the arrival time is stationary. If 
the arrival time is locally minimum, i.e., 
$(1-\kappa_i)^2-\gamma_i^2>0$ and $1-\kappa>0$,  
the density contrast satisfies 
\BE
\delta_i^{\mu}(\textrm{minima})> \frac{(2-\sqrt{2})\gamma_i \delta \kappa}
{(1-\kappa_i)^2-\gamma_i^2}, 
\EE
since $1-\kappa_i>\gamma_i$.
As one can always choose local coordinates in which $\gamma_i>0$, 
we have $\delta_i^\mu>0$. If the arrival time is locally maximum, i.e., 
$(1-\kappa_i)^2-\gamma_i^2>0$ and $1-\kappa<0$, we have  
\BE
\delta_i^{\mu}(\textrm{maxima})< -\frac{(2-\sqrt{2})\gamma_i \delta \kappa}
{(1-\kappa_i)^2-\gamma_i^2}, 
\EE
since $1-\kappa_i<\gamma_i$, leading to $\delta_i^\mu<0$.
Thus \textit{strongly lensed images
generated at a locally minimum/maximum point are magnified/demagnified 
definitely by intervening massive halos}. If a stationary point in 
the arrival time surface is a saddle one, i.e., 
$(1-\kappa_i)^2-\gamma_i^2<0$,  we have $1-\kappa_i<\gamma_i$
and $1-\kappa_i>-\gamma_i$ as we assume $\gamma_i>0$. Then 
the magnification contrast satisfies
\BE
\frac{(2-\sqrt{2})\gamma_i \delta \kappa}
{(1-\kappa_i)^2-\gamma_i^2}
< \delta_i^{\mu}(\textrm{saddle})
< -\frac{(2-\sqrt{2})\gamma_i \delta \kappa}
{(1-\kappa_i)^2-\gamma_i^2}. 
\EE
Therefore, the sign of magnification contrast 
cannot be determined definitely
without additional conditions whereas the mean value is positive 
$\langle \delta_i^\mu \rangle>0$ for $1-\kappa_i<0$
and negative $\langle \delta_i^\mu \rangle<0$ for $1-\kappa_i>0$.
If the background convergence 
satisfies a condition $2|(1-\kappa_i)|>\sqrt{2}\gamma_i$,
then the magnification contrast has a definite sign $\delta_i^\mu<0$ or
$\delta_i^\mu >0$. Thus \textit{strongly lensed images
generated at a saddle point tend to be 
demagnified/magnified by intervening massive halos if} 
$1-\kappa_i>0$$(<0)$.

It is worthwhile to note that the mass-sheet degeneracy can be
broken if intervening halos affect the fluxes of multiply
lensed images significantly. For instance, under a transformation
with a constant scalar $\lambda$ in the background convergence and shear
$1-\kappa_i \rightarrow \lambda(1-\kappa_i)=1-\kappa'_i$ 
and $\gamma_i\rightarrow 
\lambda \gamma_i=\gamma'_i$, which preserves the positions of a lensed
images of a point source by changing the position at the source plane
as $\y \rightarrow \lambda \y$.   
However, the magnification contrast $\delta^\mu_i$ depends on $\lambda$ as
$\delta_i^\mu \propto \lambda^{-1}$ for $|\delta_i^\mu| \ll 1$. Therefore, from observed
$\delta^\mu_i$, one would be able to 
put a constraint on $\lambda$ if $\delta \kappa$
and $\delta \gamma_1$ could be measured with shifts of positions of 
extended images due to intervening halos \citep{inoue2005a,inoue2005b,
vegetti2012}.

In order to quantify anomalies in flux ratios of a lens system
with quadruple images, the cusp-caustic 
relation has been used in literature. 
In the positive cusp case, for close three adjacent 
bright images (A, B and C) with magnifications
$\mu_A, \mu_B$ and $\mu_C$, and the opening angle $\theta$
spanned by the center of two images in the ends
and the lens center, the relation is   
\BE
R_{cusp}\equiv
\f{|\mu_A+\mu_B+\mu_C|}{|\mu_A|+|\mu_B|+|\mu_C|}\rightarrow 0,
\EE 
where $|\mu_A|+|\mu_B|+|\mu_C|\rightarrow \infty $ and $\theta
\rightarrow 0$. However, in practice, none of observed
quadruple image systems satisfy this asymptotic condition. For instance, 
the observed smallest 
opening angle is $\Delta \theta \sim 30^\circ$. Most radio or MIR quadruple image
systems have even larger opening angles $\Delta \theta
\sim 100^\circ$. 
Furthermore, the potential of the primary lens is sometimes not smooth.    
For example, luminous dwarf galaxies, or groups of galaxies in the
neighborhood of quadruple images can significantly alter the flux ratios. In fact,
some lens systems with anomalies in the flux ratios 
may consist of multiple lenses. Although $R_{cusp}$ is suitable for 
ideal systems with $|\mu_A|+|\mu_B|+|\mu_C|\rightarrow \infty$ 
and $\Delta \theta \sim 0$, it may not be  
suitable for most of observed quadruple-image systems.

To circumvent this problem, 
we introduce a new estimator 
\BE
\eta^2 \equiv  \f{1}{2 N_c}\sum_{i \ne j}
\biggl[ \delta_i^\mu-\delta_j^\mu(\textrm{saddle},\kappa_j<1)
 \biggr]^2,
\label{eta}
\EE
where $\delta_i^\mu$ denotes a magnification contrast for an 
image $i$ with a positive parity or a negative parity with
$\kappa_i>1$ and $\delta_j^\mu(\textrm{saddle},\kappa_j<1)$ 
is a magnification contrast for an image $j$ 
that has a negative parity with $\kappa_j<1$. Here $N_c$
is the total number of combination $i \ne j$ in the summation. 
Magnification for an unperturbed system is given by a best-fit model based on
positions of images and the center of the primary lens galaxy. 
Therefore, $\eta$ is not a directly observable quantity. 
Roughly speaking, $\eta$ corresponds to a mean magnification
contrast per image due to clustering halos in the line-of-sight.  
Note that $\eta$ depends on only 
observed and modeled flux ratios 
provided that the magnification perturbations are sufficiently 
small.  Here we put a negative sign before $\delta^\mu_j
(\textrm{saddle}, \kappa_j<1)$ because for systems with significant  
contribution from intervening halos, we expect demagnification
for saddle points with $\kappa<1$, namely,  $\langle
\delta_j^\mu\rangle<0 $ as we have seen. 
Suppose we have a set of images 
with two minima, A and C and one saddle B with $\kappa_B<1 $. 
Then the estimator of flux-ratio anomalies can be
written as 
\BE
\eta^2(\textrm{A,B,C})= \f{1}{4}[(\delta^\mu _A-\delta^\mu_B)^2
+(\delta^\mu_C-\delta^\mu_B)^2].
\label{eq:eta}
\EE 
In terms of observed fluxes $A,B,C$ and estimated unperturbed fluxes
$A_0,B_0,C_0$, the estimator is approximately given by
observed flux ratios,
\BE
\eta^2 \approx \f{1}{4}\biggl[\biggl( \f{A B_0}{A_0 B}-1 \biggr)^2
+\biggl( \f{C B_0}{C_0 B}-1 \biggr)^2 \biggr].
\EE
In a similar manner, for four-image system with 
two minima A and C and two saddles B and D with
$\kappa<1$,
the estimator is 
\BEA
\eta^2(\textrm{A,B,C,D})&=& \f{1}{8}[(\delta^\mu _A-\delta^\mu_B)^2
+(\delta^\mu_C-\delta^\mu_B)^2
\nonumber
\\
&+&(\delta^\mu _A-\delta^\mu_D)^2
+(\delta^\mu_C-\delta^\mu_D)^2  ].
\EEA

\section{Constrained convergence power }
We consider strong lens systems 
in which positions of multiple images are well fit by 
the potential of the primary lens that consists of either a single or 
multiple lenses with a smooth potential though the 
image flux ratios do not necessarily agree with the model prediction. 
In these systems, perturbation of image shifts due to other massive halos or
voids in the neighborhoods of lensed images should be sufficiently small. 
This can be interpreted as an observational selection bias that no other dark massive
halos or voids do not reside in the neighborhoods of line-of-sight of images since
presence of these objects would otherwise perturb the positions of
images and lenses significantly. Moreover, modeling the primary lens galaxies 
and neighboring groups or clusters would also induce an observational selection bias. 
The two-point correlation of matter density field in the
line-of-sight constrained by these selection biases are determined by 
the best-fitting accuracy in position of images and lenses as follows.

Suppose that positions 
of a pair of multiple images A and B with angular coordinates
$\theta_A$ and $\theta_B$ separated by an angle 
$\theta_{AB}$ are fit by a smooth lens model within an error 
$\epsilon $. This implies that total angular shifts $\delta \theta $
of image A and B due to intervening halos or voids 
in the line-of-sight 
should satisfy (see Fig.1)
\BE
|\delta \theta(\theta_A)-\delta \theta (\theta_A+\theta_{AB})|<\epsilon.
\label{eq:eps1}
\EE
Assuming statistical isotropy and homogeneity for background 
perturbations, equation (\ref{eq:eps1}) gives
\BE
2 \langle \delta\theta^2 (0) \rangle - 
2 \langle \delta\theta (0)\delta \theta(\theta_{AB}) \rangle 
<\epsilon^2, 
\label{eq:epsilon}
\EE
where $\langle \rangle $ denotes an ensemble average. 

Using Limber's approximation, a 2-point correlation 
function of astrometric shifts $\delta \theta $ 
for a pair of light rays separated
by an angle $\theta$ can be written in terms of
matter power spectrum $P_\delta(k;r)$, comoving
distance to the source $r_S$, redshift $z(r)$ as \citep{bartelmann2001}   
\BEA
\xi_{\delta \theta}(\theta) &\equiv&
\langle \delta \theta (0) \delta \theta (\theta) \rangle 
\nonumber
\\
&=&
\frac{9 H_0^4 \Omega_{m,0}^2}{ c^4}
\int_0^{r_S} dr  \biggl(\frac{r-r_S}{r_S} \biggr)^2 [1+z(r)]^2
\nonumber
\\
&\times& \int_0^{\infty}\frac{dk}{2 \pi k}  
W(k;k_{cut}(r)) P_{\delta}(k;r) J_0(g(r) k\theta),
\nonumber
\\
\label{eq:shiftsq}
\EEA
where
\BE
g(r)= \left\{ 
\begin{array}{ll}
r, & \mbox{$r<r_L$} \\
\f{r_L(r_S-r)}{r_S-r_L}, & \mbox{$r\ge r_L$}
\end{array}
\right. 
\label{eq:g}
\EE
describes the trajectory of photons that pass through
a primary lens at comoving distance $r=r_L$, $J_0$ is the zeroth-order
Bessel function, and $W(k;k_{cut}(r))$ denotes the window function 
in which modes with wavenumber $k$ smaller than 
$k_{cut}(r)$ at comoving distance $r-dr/2<r<r+dr/2$ are 
significantly suppressed. We will discuss the
property of $W(k;k_{cut}(r))$ in detail in next section. 
From equations (\ref{eq:epsilon}), (\ref{eq:shiftsq}), and (\ref{eq:g}), 
one obtains the cutoff scale $k_{cut}$ as a function of $\epsilon$ and $r$. 
Because the accuracy in position fitting is generally far better than 
that of flux ratios in observations, it may still allow deviation 
in flux ratios due to constrained convergence and shear fields in 
the line-of-sight. The constrained 2-point correlation of convergence
$\kappa$ as a function of a separation angle $\theta$ is 
\BEA
\xi_\kappa(\theta)
&\equiv&
\langle \delta \kappa (0) \delta \kappa (\theta) \rangle 
\nonumber
\\
&=&\frac{9 H_0^4 \Omega_{m,0}^2}{4 c^4}
\int_0^{r_S} dr  r^2 \biggl(\frac{r-r_S}{r_S} \biggr)^2 [1+z(r)]^2
\nonumber
\\ 
&\times &\int_0^{\infty}\frac{dk}{2 \pi} k  W(k;k_{cut}(r;\epsilon))  P_{\delta}(k;r) 
J_0(g(r) k\theta),
\nonumber
\\
\label{eq:cp}
\EEA
where $g(r)$ is given by (\ref{eq:g}). At small angular scales 
$l\gg 1$, the 
constrained convergence power is given by a Hankel transform  
of equation (\ref{eq:cp})
\BE
P_\kappa(l)=2 \pi \int d \theta \, \theta
\xi_\kappa(\theta) J_0(\theta l).
\EE
In terms of obtained constrained 
convergence correlation function $\xi_\kappa$,
one can estimate an ensemble average of the 
estimator $\eta^2$ defined in the previous section, which
measures anomaly in flux ratios.  
For example, for three images with two minima A and C and 
one saddle B with $\kappa_B<1$, using an 
approximation (\ref{eq:delta2}), for $|\delta_i^\mu|\ll1$, an ensemble average of the estimator (\ref{eq:eta}) can be written as 
\BEA
\langle \eta^2 \rangle&=&\frac{1}{4}
\biggl[(J_A+J_B)\sigma_\kappa^2(0) 
-2J_{AB}\xi_\kappa (\theta_{AB}) 
\label{eq:estimator-anal}
\nonumber
\\
&+&(J_B+J_C)\sigma_\kappa^2(0) 
-2J_{BC}\xi_\kappa(\theta_{\textrm{BC}}) \biggr],
\EEA
where
\BE
J_i=\mu_i^2(4(1-\kappa_i)^2+2 \gamma_i^2),
\label{eq:Ji}
\EE
and
\BE
J_{ij}=\mu_i \mu_j(4(1-\kappa_i)(1-\kappa_j)+2\gamma_i \gamma_j),
\label{eq:Jij}
\EE
for $i=A,B,C$ and $\gamma_i=(\gamma_{i1}^2+\gamma_{i2}^2)^{1/2}$.  
Here $\sigma_\kappa(0) \equiv \sqrt{\xi_\kappa(0)}$.
In deriving equation (30), we have used a well known fact that 
$\xi_\kappa(\theta)=2\xi_{\gamma_\alpha}(\theta)$
and $\langle \delta\gamma_1 \delta\gamma_2 \rangle= \langle \delta
\kappa \delta \gamma_\alpha  \rangle=0 $, for $\alpha=1,2$
provided that background matter density 
fluctuations are statistically homogeneous and isotropic.
In a similar manner, for a four-image system with 
two minima A and C and two saddles B and D with
$\kappa<1$,
an ensemble average of the 
estimator $\eta^2(\textrm{A,B,C,D})$
is given by
\BEA
\langle \eta^2 \rangle&=&\frac{1}{8}
\biggl[(J_A+J_B)\sigma_\kappa^2(0) 
-2J_{AB}\xi_\kappa(\theta_{AB}) 
\nonumber
\\
&+&(J_C+J_B)\sigma_\kappa^2(0) 
-2J_{CB}\xi_\kappa(\theta_{\textrm{CB}}) 
\nonumber
\\
&+&(J_A+J_D)\sigma_\kappa^2(0) 
-2J_{AD}\xi_\kappa(\theta_{\textrm{AD}}) \nonumber
\\
&+&(J_C+J_D)\sigma_\kappa^2(0) 
-2J_{CD}\xi_\kappa(\theta_{\textrm{CD}}) \biggr],
\EEA
where $J_i$ and $J_{ij}$ are given by (\ref{eq:Ji}) and (\ref{eq:Jij}).
Application of these statistics to mid-infrared lenses is presented in section 6.

\section{Non-linear power spectrum}

In order to evaluate equations (\ref{eq:shiftsq}) and (\ref{eq:cp}),  
we need an accurate matter power spectrum at scales down to
 $k^{-1} =O[1]\,h^{-1}$ kpc.
However, analytical fitting formulae in literature are not
suitable for this purpose.
For example, the halo-fit model by \citet{smith2003} has been frequently
 used to evaluate the non-linear power spectrum $P(k)$.
 However, as shown by several authors (see e.g. \citet{takahashi2012}) 
it was shown that this model underestimates the power spectrum by
 some tens percent than the latest cosmological simulation results
 on small scales $k \gtrsim 1\,h {\rm Mpc}^{-1}$.
Hence, in this study, we run cosmological $N$-body simulations to investigate
 the non-linear power spectrum $P(k)$ at 
galactic scales and we make a new fitting formula of $P(k)$.
For simplicity, however, we do not input baryon in our simulation. 

In our cosmological $N$-body simulation, we use 
a cubic box with a comoving side length of $10\,h^{-1}\,$Mpc
with $N_p^3=1024^3$ and $512^3$ collisionless particles. 
We can check a numerical convergence of our simulation by comparing
a low resolution simulation ($N_p^3=512^3$) with 
a high resolution one ($1024^3$).
The softening comoving length is fixed to be $2.5\%$ of the mean particle separation,
corresponding to $0.25 (0.5)\,h^{-1}\,$kpc for $N_p^3=1024^3 (512^3)$.
The particle mass is $7.1 \times 10^4 (5.7 \times 10^5)\, h^{-1} M_\odot$
 for  $1024^3 (512^3)$ collisionless particles. Therefore, the minimum
mass of halos resolved by our simulations is $1.4 \times
10^5\,h^{-1}\ms$ (corresponding to 20 particles).

We use a simulation code
 called Gadget2 \citep{springel2001,springel2005}. 
We calculate the initial conditions of particles 
based on the second-order Lagrangian perturbation theory (2LPT)
\citep{crocce2006,nishimichi2009} with
the initial linear power spectrum obtained by 
\citet{eisenstein1999}.
The initial redshift of our simulations 
is $z_{\rm in}=99$ and we dump the simulation
results of the particle positions at $z=0-4$.
We prepare two independent realizations for $N_p^3=512^3$ at 
$z=0,0.35,0.7,1,1.5,2.2,3,4$ and a single
 realization for $N_p^3=1024^3$ at $z=0.35,0.7,1,1.5,2.2,3,4$.
In calculating the power spectrum $P(k)$, we assign the particles 
on $N_g^3=1280^3$ grid using 
the cloud-in-cell (CIC) method to obtain density fluctuations.
After performing the Fourier transform,
 we correct the window function of CIC as $\tilde{\delta}_{\bf k}
 \rightarrow \prod_{i=x,y,z} \left[ {\rm sinc} (L k_i/2 N_{\rm g}) \right]^2 \times
 \tilde{\delta}_{\bf k}$, where $\tilde{\delta}_{\bf k}$ is the density
 fluctuation in the Fourier space (Hockney \& Eastwood 1988).
In addition, to evaluate the power spectrum on small scales accurately,
 we fold the particle positions ${\bf x}$ into a smaller box by replacing  
 ${\bf x} \rightarrow {\bf x} \% (L/2^n)$ where the operation $a \% b$ gives
 the reminder of the division of $a$ by $b$ (e.g. \citet{valageas2011}).
Then, the resolution becomes effectively $2^n$ times finer.
Here we use $n=2$ and $4$.
Finally, we evaluate the power spectrum:
\BE
  P(k) = \sum_{k} \frac{1}{N_k} \left| \tilde{\delta}_{\bf k} \right|^2,
\EE
where the summation is done from $k-\Delta k/2$ to $k+\Delta k/2$ with
 a binwidth $\Delta k$, and $N_k$ is the number of mode in the bin.
We use a logarithmic binwidth, $\Delta \log_{10} ( k/h{\rm Mpc}^{-1})=0.05$.
For our polynomial fitting, we do not use the $P(k)$ at small scales 
where the shot noise begin to dominate the signal.
The Nyquist wave number determined by the mean particle separation is
 $k_{\rm Nyq}=(2 \pi/L) (N_p/2)$, which corresponds to $k_{\rm Nyq}
 =320 (160)\, h {\rm Mpc}^{-1}$ for $N_p^3=1024^3 (512^3)$ particles
 with $L=10h^{-1}$Mpc. 
Hence, we can probe the density fluctuations on very small scales,
 $k=320\,h {\rm Mpc}^{-1}$.
We have checked that the power spectra $P(k)$ of our $N$-body simulations
 agree with simulations with higher resolution in which we use finer
 simulation parameters of the time steps, force calculation, etc.
 within $2(6\%)$ for $k<100(320)\,h {\rm Mpc}^{-1}$.

We also use simulation results by \citet{takahashi2012}.
They use the same codes as ours (Gadget2 and 2LTP initial condition)
and employ $1024^3$ particles in the simulation boxes of $L=2000,800,
 320\,h^{-1}$ Mpc and combined the $P(k)$ on the 
different box sizes to cover a wide range of scales.
They provide the $P(k)$ up to $k=30\,h \textrm{Mpc}^{-1}$ at $z=0-10$.
Hence we use their result for $k<30\,h \textrm{Mpc}^{-1} $ and the present 
result for $k>30\,h \textrm{Mpc}^{-1}$. 

\begin{figure}
\vspace*{1.0cm}
\includegraphics[width=85mm]{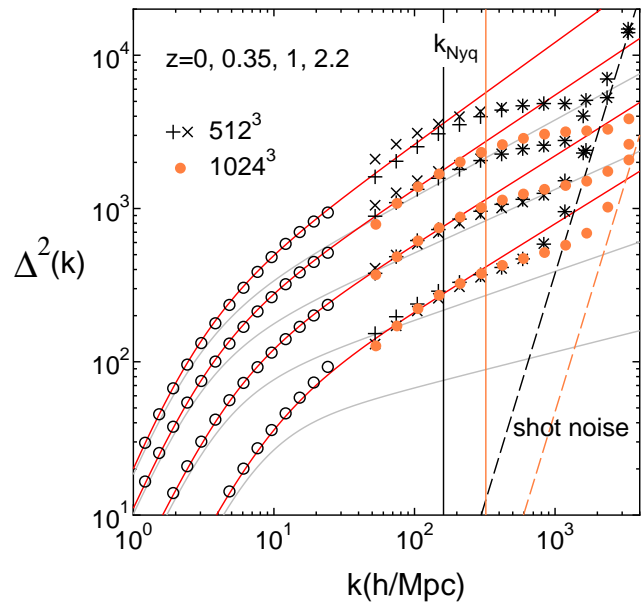}
\caption{
Plots of dimensionless power spectra, $\Delta^2(k)= k^3 P(k)/(2 \pi^2)$,
 at $z=0,0.35,1$ and $2.2$ (from top to bottom). Our simulation 
results of $L=10\,h^{-1}$Mpc box 
 with $512^3$ particles are plotted as plus and cross symbols,
corresponding to two independent realizations.
Our simulation results of $L=10\,h^{-1}$Mpc box 
with $1024^3$ particles and 
those by \citet{takahashi2012}, which match our simulation results 
at $k=30\,h{\rm Mpc}^{-1}$ are plotted as filled circles and empty circles, respectively.
Our fitting functions (see Appendix) and the predicted values in  
the halo-fit model \citep{smith2003}
are plotted as solid curves in red and gray, respectively. 
The vertical and dashed lines represent the Nyquist wave numbers
 $k_{\rm Nyq}$ and the shot noises for $N_p^3=1024^3$ (right) and $512^3$
 (left), respectively.
}
\label{f2}
\vspace*{0.2cm}
\end{figure}

As one can clearly see in Fig. \ref{f2},
the power spectra of our simulations at scales $k>30\,h{\rm Mpc}^{-1}$ 
are significantly larger than the values predicted in 
the original halo-fit model \citep{smith2003}. At scales $k \sim
300\,h{\rm Mpc}^{-1}$ which are relevant to the weak lensing effect
of line-if-sight halos, our fitting formula, which improves the original
halo-fit model down to $k\sim 300\,h{\rm Mpc}^{-1}$ (see Appendix for
details) predicts a factor of $2-3$ enhancement in amplitude of the dimensionless
power $\Delta(k)=\sqrt{\Delta^2(k)}$ at redshifts $0<z<2$. As we shall
see in the following sections, this enhancement plays an important role
in explaining the origin of anomalies in flux ratios.

The suppression of power spectra $P(k)$ predicted by our simulation at scales smaller than
the Nyquist frequency is due to the lack of resolution.
In fact, the power spectra $P(k)$ in simulations with $1024^3$
particles are systematically larger than those simulated with
 $512^3$ particles at $k>100\,h{\rm Mpc}^{-1} $. In what follows, however, 
we use our new fitting formula to estimate the power spectra
at very small scales $k>320\,h{\rm Mpc}^{-1}$. Although the accuracy is not
guaranteed, it may give a good approximation on very small scales
since the power tends to increase as the
resolution becomes higher.

\begin{figure}
\vspace*{1.0cm}

\includegraphics[width=80mm]{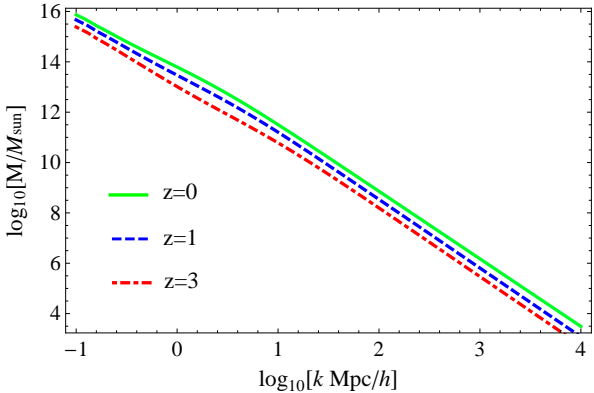}
\caption{Approximated relation between mass $M$ and a wavenumber $k$ for non-linear 
matter fluctuations at redshift $z=0$(full), $z=1$(dashed), and 
$z=3$(dot-dashed).   }
\label{f3}
\vspace*{0.2cm}
\end{figure}

Assuming that the comoving size $r$ of a density fluctuation is 
$r\sim \pi/k$, the relation between a mass scale of clustering 
non-linear halos and the wavenumber can be given by 
\BE
M(k;z)\sim \frac{4 \pi^4}{3k^3}\sqrt{\Delta^2(k) }\rho(z),
\EE
where $\rho(z)$ is a mean comoving matter 
density of the background universe at redshift $z$.
As shown in Fig. \ref{f3}, for a given wavenumber $k$, the corresponding mass
$M$ increases as the redshift $z$ increases due to time evolution of 
the fluctuations. We find that a wavenumber that corresponds to the 
Nyquist frequency $k_{\rm Nyq}=320\,h{\rm Mpc}^{-1}$ 
corresponds to $M=3\times 10^7\, \ms$ at $z=0$. 
$k=1000\,h{\rm Mpc}^{-1},~10000\,h{\rm Mpc}^{-1}$ 
corresponds to $M=1\times 10^6\ms,~ 3\times 10^3\, \ms$ at $z=0$,
respectively. 


\section{Semi-analytic estimate}
Based on formalism developed in section 2 and 3,
we can estimate the deflection and flux change 
of strongly lensed images using the 
non-linear matter power spectrum obtained in 
section 4. Firstly, we calculate the rms of 
difference in the total angular shifts $\delta \theta $ between two images
separated by $\theta$,
\BE 
\Delta \delta \theta(\theta)\equiv \bigl[\langle 
(\delta\theta(\theta)-\delta\theta(0))^2 \rangle \bigr]^{1/2}.
\EE
In order 
to study the dependence on the scale of matter 
fluctuation in the line-of-sight, for simplicity, we 
adopt a ``sharp k-space'' window function,
\BE
W_s(k;k_{cut})\equiv \left\{
\begin{array}{ll}
0\,, &  k<k_{min}\\
1\,, &  k \ge k_{min} \\
\end{array}
\right.
\label{eq:W}
\EE
for which modes $k<k_{min}=k_{cut}$ are cut off in equation (\ref{eq:shiftsq}). 
\begin{figure}
\vspace*{1.0cm}
\includegraphics[width=85mm]{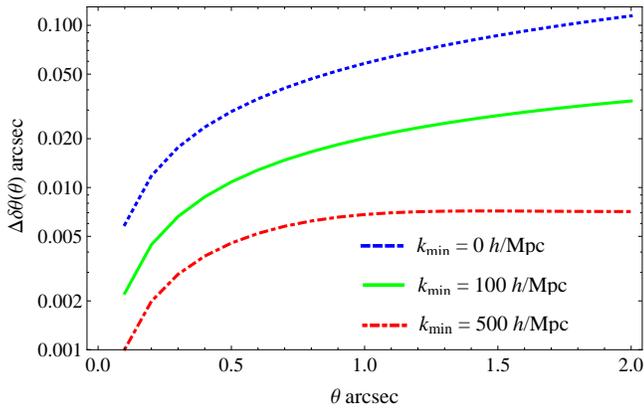}
\caption{Rms value of difference in shifts of image position  
$\Delta \delta \theta $ as a
 function of a separation angle $\theta$ for minimum wavenumber 
$k_{min} = 0\, h{\rm Mpc}^{-1}$ (dashed), $100 \,h{\rm Mpc}^{-1}$  (full), 
$500 \,h{\rm Mpc}^{-1}$ (dot-dashed). Source and lens redshifts are 
$z_S = 3, z_L = 0.5$, respectively. The upper limit of the wavenumber
is assumed to be $k_{max}=1000\, h{\rm Mpc}^{-1} $. }
\label{f4}
\vspace*{0.2cm}
\end{figure}
As shown in Fig. $\ref{f4}$, for $k_{min}<100\, h{\rm Mpc}^{-1}$, 
the difference in the image shift $\Delta
\delta \theta$ becomes larger as the separation angle increases. 
However, for $k_{min}=500\, h{\rm Mpc}^{-1}$, $\Delta \delta \theta$
has a peak around $\theta=1.5$ arcsec and it gradually 
decreases as $\theta $ increases due to a cutoff at small scales
$k>k_{max}$.
For a given separation angle, $\Delta \delta \theta$ gets smaller as $k_{min}$
increases. 

\begin{figure}
\vspace*{1.0cm}
\includegraphics[width=85mm]{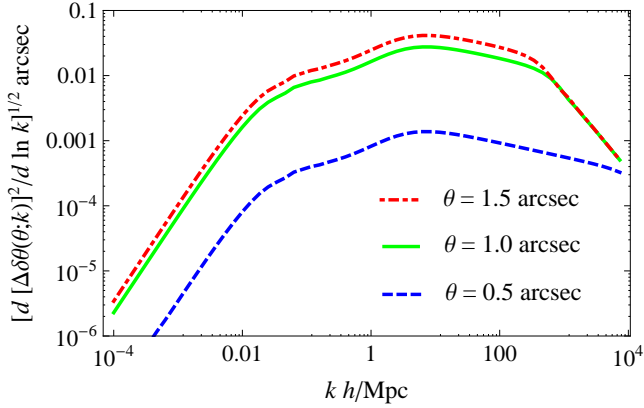}
\caption{Contribution of each mode 
to $\Delta \delta \theta$ marginalized over redshift $z$ 
as a function of wavenumber $k$.
For separation angle $\theta\sim 1$, 
major contribution to $\Delta \delta \theta$ comes from modes 
$k \sim 10\, h{\rm Mpc}^{-1}$.
Plotted curves are for separation angle 
$\theta = 0.5''$ (dashed), $1.0''$ (full), 
$1.5''$ (dot-dashed) with source redshift $z_S = 3$ and lens redshift 
$z_L = 0.5$. We assume $k_{max}=10000\,h{\rm Mpc}^{-1}$.  }
\label{f5}
\vspace*{0.2cm}
\end{figure}

\begin{figure}
\vspace*{1.0cm}
\includegraphics[width=85mm]{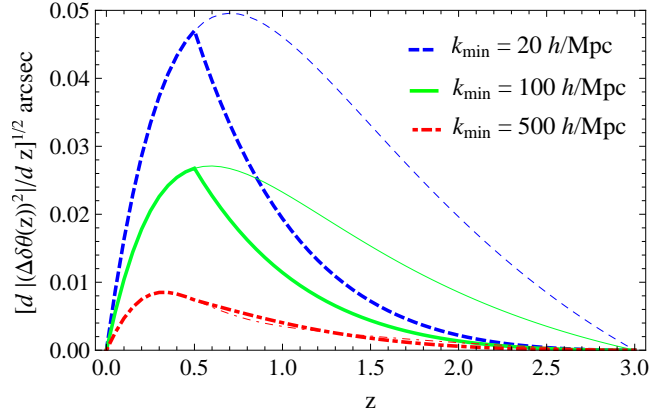}
\caption{Contribution of fluctuations on different planes 
at redshift $z$ to $\Delta \delta \theta$ marginalized over wavenumber 
$k$. Plotted curves are 
for minimum wavenumber 
$k_{min} = 20 \,h{\rm Mpc}^{-1}$ (dashed), 
$100 \,h{\rm Mpc}^{-1}$ (full), and
$500 \,h{\rm Mpc}^{-1}$ (dot-dashed).
Thick and thin curves correspond to lens redshifts
$z_L=0.5$ and 
$z_L=2.99$, respectively.
 We assume source redshift  
$z_S = 3$, separation angle $\theta=1\,$arcsec, 
and the upper limit of the 
wavenumber $k_{max}=1000\,h{\rm Mpc}^{-1}$.     }
\label{f6}
\vspace*{0.2cm}
\end{figure}

Without any cutoff in the background fluctuations, it turns out that the
difference in the shifts is
$\Delta \delta \theta \sim 0.06$ arcsec for $k_{max}=1000\,h
\textrm{Mpc}^{-1}$ and 
separation angle $\theta=1$ arcsec, which is the 
typical scale of the Einstein radius of 
a primary lens.  This is significantly larger than   
the observed error $\epsilon$ of relative positions of light centers, which is 
of the order of $1$ mas for most quadruple 
image lenses observed in the optical or radio band. 
This is because contribution from modes with wavenumber 
$k<k_{min}=O[10^{2-3}]h{\rm Mpc}^{-1}$ is already taken into account in
the fitted model, including neighboring clusters, groups, massive galaxy
halos and luminous dwarf galaxy halos provided that the 
position of modeled images and the center of the primary lens agree with the
observed values. In fact, as shown in Fig. \ref{f5} 
on angular scales smaller than $1\,$arcsec,  
the largest contribution comes from modes on scales $k\sim
10\,h{\rm Mpc}^{-1}$ at which the one-halo term begins to dominate the
matter power
spectrum in the halo fitting model. For larger separation angles
$\theta>1\,$ arcsec, relative contribution from smaller scale fluctuations
decreases. If the residual difference in the position of 
lensed images is less than 
the rms value, then contribution from clustering halos 
on such scales should be negligible. As we can see in Fig. \ref{f6}, 
if the cutoff scale $k_{min}$
is constant as a function of redshift as we have assumed, 
then the dominant contribution comes from fluctuations near
the lens plane at $z \sim z_L$ in the line-of-sight. 
This can be understood as a result of two effects :(i)
The lensing weight function $(r-r_S)(1+z(r))$ for astrometric
shifts in the integrand of equation (24) is maximum at 
around a half comoving distance to the source,
and vanishes at $r=r_S$. (ii) Because of 
convergence of photon traj
ectories toward a point-like source 
beyond the lens plane, the relevant comoving scale 
of fluctuations becomes smaller, which leads to a suppression 
of the contribution. The latter effect (ii) becomes more 
significant as a cutoff scale becomes larger and the lens redshift
$z_L$ becomes smaller.  

Thus we interpret that the contribution from fluctuations on scale 
$k\sim 10\,h{\rm Mpc}^{-1}$ at $\sim z_L$ correspond to 
a primary lens galaxy or luminous galaxies around the 
primary lens. If large-angle fluctuations of projected gravitational potential 
in the line-of-sight of multiply lensed images have been modeled 
as an external shear ($m=2$) or low multipole terms ($m=3$), 
which corresponds to contribution from clusters or groups of galaxies, 
we would only need to take into account small scale fluctuations 
with wavenumbers $k \gtrsim k_{cut}$ where $k_{cut}$ is roughly equivalent to
the scale of observable luminous dwarf galaxies at the lens plane. 
These small scale fluctuations  
may affect the flux ratios significantly if the 
total perturbations of convergence integrated in
the line-of-sight is sufficiently large.

\begin{figure}
\vspace*{1.0cm}
\includegraphics[width=85mm]{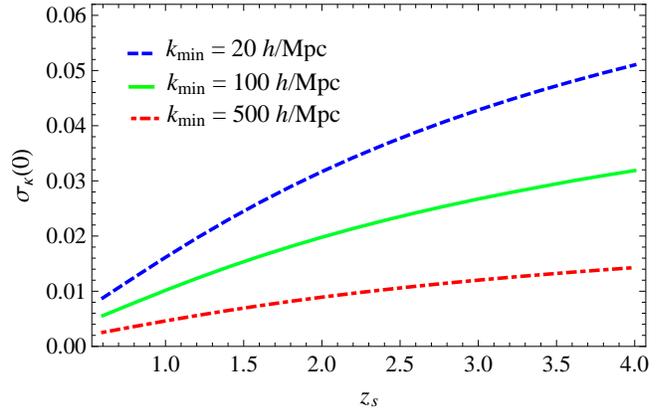}
\caption{Self correlation of convergence as a function of source 
redshift $z_S$. Plotted curves are for minimum wavenumber 
$k_{min} = 20\, h{\rm Mpc}^{-1}$ (dashed), 
$100 \,h{\rm Mpc}^{-1}$ (full), and
$500 \,h{\rm Mpc}^{-1}$ (dot-dashed). We assume lens redshift  
is $z_L = 0.5$ and the upper limit of the 
wavenumber is $k_{max}=1000\,h{\rm Mpc}^{-1}$.    }
\label{f7}
\vspace*{0.5cm}
\end{figure}

\begin{figure}
\vspace*{1.0cm}
\includegraphics[width=85mm]{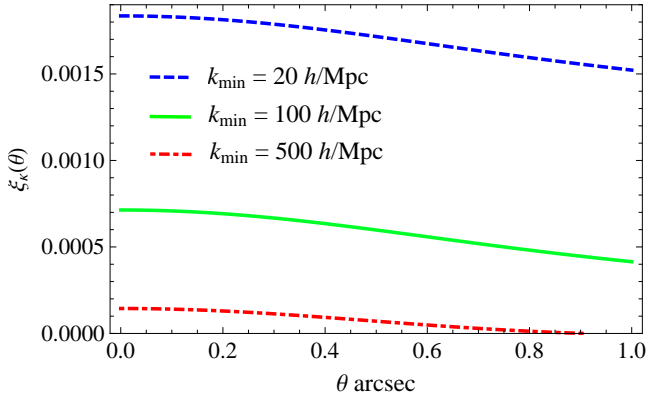}
\caption{2-point correlation of convergence as a function of separation
angle $\theta$ between two images. The three curves are for
$k_{min} = 20 \,h{\rm Mpc}^{-1}$(dashed), $100\, h{\rm Mpc}^{-1}$ (full), 
and $500 \,h{\rm Mpc}^{-1}$ (dot-dashed).  We assume lens redshift
 $z_L=0.5$, source redshift  
$z_S = 3$, and the upper limit of the 
wavenumber $k_{max}=1000\,h{\rm Mpc}^{-1}$. }
\label{f8}
\vspace*{0.2cm}
\end{figure}

In order to assess the effects of small scale fluctuations 
on the flux ratios, we consider
rms of the self correlation of convergence $\sigma_\kappa(0)=\sqrt{\xi_\kappa(0)}$
as a function of source redshift $z_S$.
We can see in Fig. \ref{f7} that contribution of small scale 
fluctuations ranging from $k_{min}\sim O(10^2)h{\rm Mpc}^{-1}$
to $k_{max}=1000\, h{\rm Mpc}^{-1}$ would 
yields $\sigma_{\kappa}(0)\sim 0.01$ in convergence if the source redshift 
satisfies $z_S \gtrsim 3$\footnote{If we consider contribution from
modes with wavenumber $k>1000 \, h{\rm Mpc}^{-1}$, the rms convergence $\sigma_\kappa(0)$ can
be further increased. See also Fig. 9.}. In other words, the surface 
density in small scale structures in line-of-sight is of the order of 
one percent of that of the primary lens. This may be too 
small to be of any importance. However, as we have seen in section 3, 
\textit {anomaly in the flux ratio 
is proportional to the magnification of the primary lens}, i.e., 
$\eta \sim 2 \langle \mu \rangle  \sigma_{\kappa}(0)$ where $\mu$ is the
mean magnification of images provided that $\kappa\sim \gamma\sim
0.5$ and correlations between different images are negligible.  
For an image with modest magnification $\mu \sim 5$, it would
yield 10 per cent change in flux ratios, $\eta \sim
0.1$ if $z_S \sim 3$. Such a change is sufficient to explain the 
order of observed anomalies \citep{metcalf2005a, metcalf2005b}. 
Moreover, if we take into account
the correlation of convergence 
between different images, anomaly in flux ratios can be more distinctive. 
As we can see in Fig. \ref{f8}, 
for a separation angle $\theta \sim 0.5\,$ arcsec, 
the amplitude of 2-point correlation $\xi_\kappa(\theta)$ is still
comparable to the self correlation $\sigma_\kappa^2 (0)$. 
Therefore, we expect less significant anomaly in the flux ratios
for systems with larger Einstein radius as long as accuracy in position 
fitting does not change. In other words, if such an anomaly is observed
in the primary lens with large separation angles, the chance of 
significant perturbation to one of the lensed 
images is higher than the cases in which all the images are
perturbed at the same time at similar levels.

\begin{figure}
\vspace*{1.0cm}
\includegraphics[width=85mm]{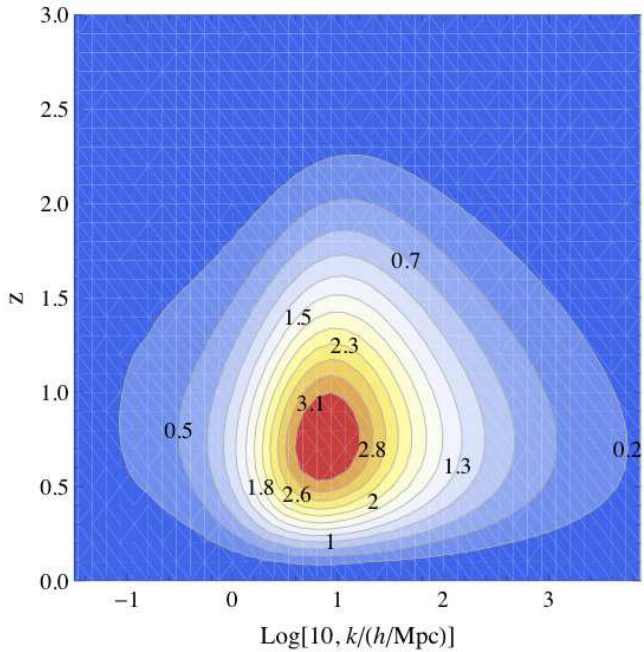}
\caption{Contour plots of $\del^2 \sigma^2_\kappa(0)/\del z \del
 \ln{k}$.  We assume lens redshift $z_L=0.5$, source redshift  
$z_S = 3$, and the upper limit of the 
wavenumber $k_{max}=10000\,h{\rm Mpc}^{-1}.$
}
\label{f9}
\vspace*{0.2cm}
\end{figure}

As is the case of shifts of image positions,
the largest contribution to the amplitude of convergence $\sigma_\kappa(0)$ 
comes from modes on scales $k \sim 10 \, h{\rm Mpc}^{-1}$
at approximately a half distance to the source (Fig. \ref{f9}).
If we consider a sharp k-space filter with $k_{cut}=k_{min}$,
then the small scale contribution to the convergence $\sigma_\kappa(0)$ 
is smaller than contribution from modes with $k\sim k_{min}$. 
In other words, the largest contribution to the amplitude of 
convergence $\sigma_\kappa(0)$ comes from fluctuations with wavenumber $k \sim k_{min}$
at approximately a half distance to the source. Therefore, we consider 
that influence of redshift dependence of filtering functions
$W(k;k_{cut}(z))$ on $\sigma_{\kappa}(0)$ is small.

\begin{figure}
\vspace*{1.0cm}
\includegraphics[width=85mm]{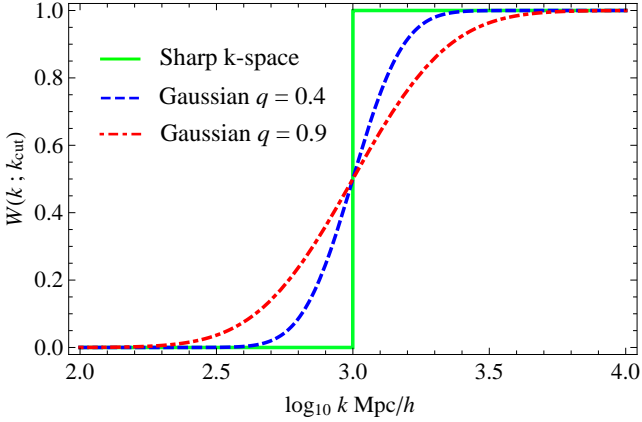}
\caption{The ``Gaussian'' filters with $q=0.4$ (dashed curve)
$q=0.9$ (dot-dashed curve) and the 
sharp k-space filter (full curve). We assume 
$k_{cut}=1000\,h{\rm Mpc}^{-1}$. }
\label{f10}
\vspace*{0.2cm}
\end{figure}

So far we have considered the ``sharp k-space'' filter
for cutting off the contribution from large scale 
fluctuations with $k<k_{min}$. However, in real setting,  large scale 
fluctuations in the line-of-sight are removed 
in real space as components that yield an external shear 
and a constant convergence or distort the relative positions of
the lensed images and the lens centroid. 
Therefore, contributions from modes $k<k_{min}$ may
not be negligible. In other words, more massive 
objects placed near the line-of-sight could affect the
flux ratios.
In order to assess this effect,
we consider a ``Gaussian'' filter that is defined as 
an integration of the Gaussian distribution function
as
\BE
W_g(k;k_{cut})\equiv\frac{1}{2}\biggl[1+\textrm{Erf}
\biggl( 
\frac{\log_{10}(k/k_{cut})}{\sqrt{2} \log_{10}(1+q) }
\biggr) 
\biggr],
\EE   
where Erf is the error function and $q$ describes the width of the filter ${\it{\Delta}}
\log_{10}k\sim q $. There should be an upper limit for $q$
as the perturber in the line-of-sight is 
too massive, it becomes observable near the primary lens. 
As a reasonable guess, we consider two types of the ``Gaussian''
filter, $q=0.4$ and $q=0.9$. Approximately 4 times 
massive objects are included for $q=0.4$ and
20 times massive for $q=0.9$ (Fig. \ref{f10}). 
We choose two types of ``UV'' cutoff,
$k_{max}=1000\, h{\rm Mpc}^{-1}$ and 
$k_{max}=10000\, h{\rm Mpc}^{-1}$.
The latter scale gives  
an Einstein radius $O[1\,\textrm{pc}]$ at cosmological scales 
if the fluctuations of corresponding mass scales
$\sim 10^3 \ms$ form point masses. 

\begin{figure}
\includegraphics[width=85mm]{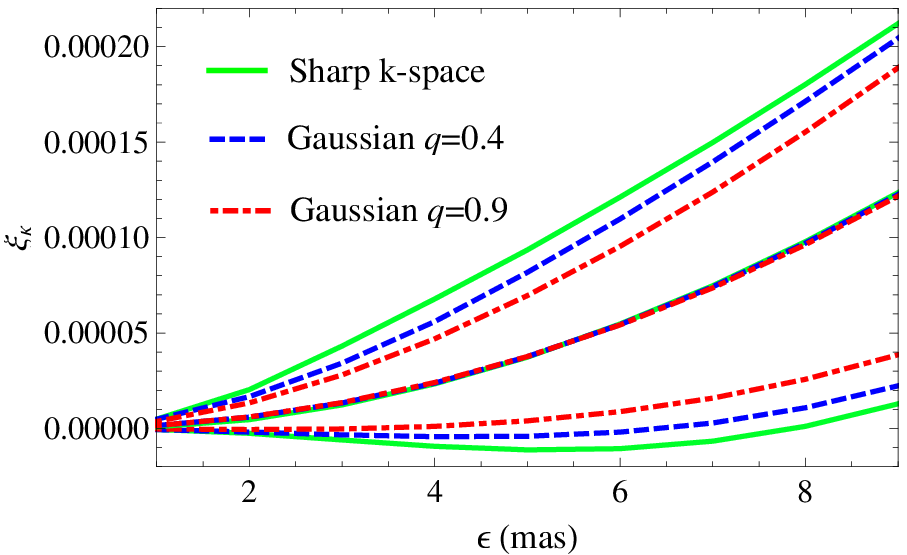}
\includegraphics[width=85mm]{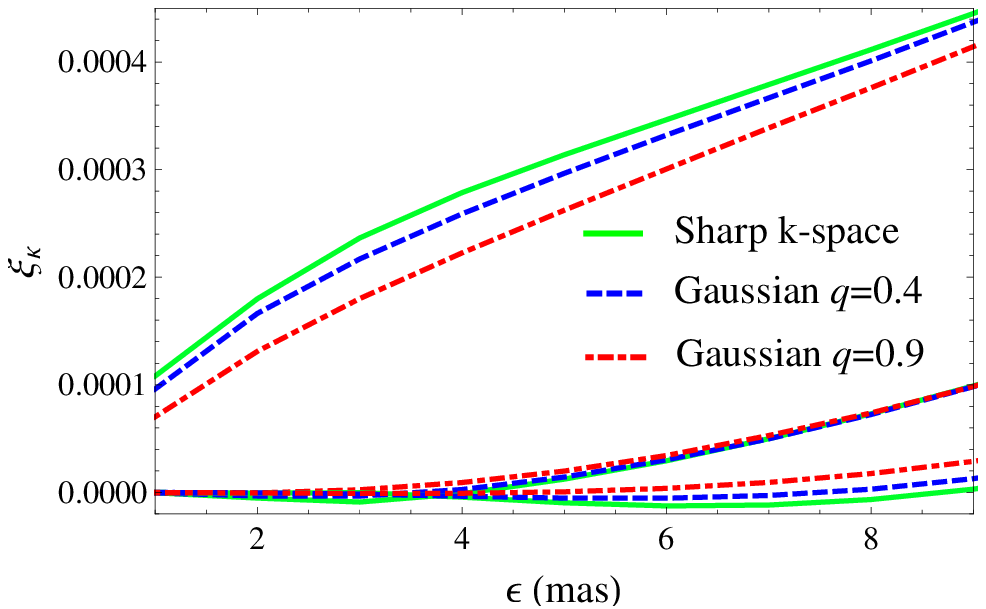}
\caption{2-point correlation functions of convergence 
$\xi_\kappa(\theta)$ for
the ``Gaussian'' filters with $q=0.4$ (dashed curves),
$q=0.9$ (dot-dashed curve) and for the sharp k-space filter (full curve)
as a function of the rms shift $\epsilon$ 
in the relative position of a pair of images
separated by $\theta$.  The upper and the lower panels show 
the plots for $k_{max}=1000\, h{\rm Mpc}^{-1}$ and for 
$k_{max}=10000\, h{\rm Mpc}^{-1} $, respectively. In each panel,
the top, the middle, and the bottom three curves correspond
to $\theta=0''$, $\theta=0.5''$, and $\theta=1''$, respectively. 
We assume $z_S=3$ and $z_L=0.5$.  }
\label{f11}
\vspace*{0.2cm}
\end{figure}

For a given maximally allowed 
shift $\epsilon=\Delta \delta \theta(\theta=1'')$,
$k_{cut}$ is obtained from equations (24) and (25) for each filter. Then
we calculate the 2-point correlation function of convergence $\xi_{\kappa}$.
We adopt separation angles between an image and a lens
$\theta=0'', 0.5'',$ and $1''$ as typical examples.  
As shown in Fig. \ref{f11}, the differences between different types of filters
are astonishingly small for $\theta=0.5''$. For $\theta=0, 1''$, the
relative difference in $\xi_\kappa$ is at most 35 per cent.   
Therefore, the effect of large scale fluctuations
with $k<k_{cut}$ to $\eta$ is less than $\sim 20$ per cent. 
It should be noted, however, that for $\theta=0''$, amplitude of 
$\xi_\kappa$ is systematically reduced
if contribution from large scale fluctuations is taken into account. 
This can be explained as follows.
At $\theta \sim 0''$, the ratio between contributions to the shift $\Delta
\delta$ from large scales $k<k_{cut}$ and small scales
$k>k_{cut}$ is {\textit{smaller}} than the ratio between contributions
to the convergence $\xi_{\kappa}$ from 
large scales $k<k_{cut}$ and small scales $k>k_{cut}$.
In other words, the contributions to the 2-point correlation of 
convergence is a steeper function of $k$ in comparison with the 
contribution to the shift $\Delta \delta \theta $ (see Fig.5 and Fig. 9).   
As the function form of the filter function is common
for the both quantities, the above relation yields a further reduction
in the 2-point correlation $\xi_{\kappa}$ if the cut off scale is determined
from the shift $\Delta \delta \theta$. 
At $\theta=1''$, on the other hand,  
the ratio between contributions to the shift $\Delta
\delta$ from large scales $k<k_{cut}$ and small scales
$k>k_{cut}$ is {\textit{larger}} than the ratio between contributions
to the convergence $\xi_{\kappa}$ from 
large scales $k<k_{cut}$ and small scales $k>k_{cut}$.
This yields an enhancement of 2-point correlation function $\xi_{\kappa}$.
If $q>0.9$,
we expect that the predicted $\eta$ will be much smaller.  

\section{MIR QSO-galaxy quadruple lenses}

In what follows, we only use MIR data for
flux of gravitationally lensed QSOs with quadruple images (see table 1).
The number of lensed images used in analysis of flux ratios is 
denoted as $N$. We discard any data with small 
signal-to-noise ratio in the MIR fluxes. 
Relative positions of macro-lensed 
images and the lensing galaxy are taken from the 
CfA-Arizona Space Telescope Lens Survey (CASTLES)\footnote{See
http://www.cfa.harvard.edu/castles/}
in the visible and near-IR bands except for H1413+117.

For four-image lenses, the mean 
error in the image separation between a lensed image
and a position of the centroid of the primary lens  
is denoted as $\langle \epsilon \rangle$. We assume that 
the errors of image positions or a centroid of lensing galaxy are not
correlated each other and the errors in image 
separation obey Gaussian distributions. 
Although constraint on contribution from intervening halos
is most stringent for a pair of lensed 
images with the largest separation angle,
as a simple approximation, we adopt a mean separation angle
$\langle \theta \rangle$ between a lensed image and a position of a
centroid of the primary lens obtained from all 4 images and their 
errors in positions as observable quantities that can be used for 
constraining intervening halos from astrometric shifts\footnote{In more
realistic setteing,
we consider light rays that pass through the lensed 4 images (2 saddle
and 2 minima) and 1 maximum at the lens plane. Since the position of 
the maximum is usually close to the centroid of the lensing galaxy, we
assume that the corresponding light ray is approximated by a geodesic 
that pass through the centroid of lensing galaxy.}.    
This is because any contribution from 
fluctuations with angular size similar to the largest separation angle 
can be taken into account as a part of the constant background 
convergence and shear. Fluctuations with angular size
similar to $\langle \theta \rangle$ does not contribute to the 
constant background convergence and shear but they may change the 
separation angle between a lensed image and the source. 
In what follows, we also assume that the centroids of MIR
images agree with those observed in the visible and near-IR bands.

In this study, we do not use radio QSO lenses, which have been frequently 
used in literature because the finite source-size effect 
might be important in analyzing flux-ratios perturbed by intervening
clustering halos. The typical size of radio continuum 
emission region of QSO lenses is $L \sim 10\, \textrm{pc}$. For lens systems with
magnifications above $\mu=10$, then the sizes of magnified images 
can be estimated as $\gtrsim 3\times 10\, \textrm{pc}$. Assuming
that the size of the Einstein radius of is typically
$x_E \sim 5 \,\textrm{kpc}$, a fractional difference in magnification 
with respect to a point-source 
is $\delta \mu/\mu \sim |\ln(L/x_E)| L/x_E\sim 0.05 $
\citep{inoue2005a} 
if a top-hat type source with an apparent size $L=50\, \textrm{pc}$
at the lens plane is placed at the center of an SIS. Even in 
more realistic cases, the order of the
difference would be the same as long as the potential has a form
similar to an SIS. For instance, the correction term due to differential
magnification is proportional to $L/x_E$ for an SIE lens because we have  
$\mu^{-1}\sim 1-O[x_E/x]$ \citep{kormann1994} where $x$ denotes the 
radial proper distance from the center of an SIE. Moreover, 
inclusion of substructure in the primary lens can boost the 
perturbation by a factor of 2-3 \citep{metcalf2012}. Therefore, we
expect $\sim 10$ per cent systematic change in the flux ratios. 

The size of continuum emission regions in the MIR band is
typically much smaller than radio counterparts. 
In fact, the estimated source sizes in our sample of MIR lenses are 
$L \sim 1\,\textrm{pc}$, which is significantly larger than the Einstein radius of 
stars $L \sim 0.01\, \textrm{pc}$. Therefore, our sample is free 
from the finite source-size effect and the microlensing effects due to stars.
For 5 lenses in our sample, the point-like
source approximation is valid at $\sim 1$ per cent
level in flux ratios as long as magnification is not significantly
large, i.e., $\mu \lesssim 10 $. Moreover, we note that the effect of differential 
magnification due to intervening halos is negligible as long as the order of shifts in the
relative position of images divided by the Einstein radius of the
lensing galaxy is $\delta x/x_E=O(0.001)$ since the magnification
perturbation due to shifts of $\delta x$ can be estimated as $\delta \mu/\mu\sim
\delta x/x_E$. 

As a fiducial model of these lenses, we adopt a singular isothermal ellipsoid 
(SIE) plus an external shear (ES) \citep{kormann1994}, which can explain
flat rotation curves. We use only
relative positions of lensed quadruple images and the center of lensing galaxy
for modeling. The parameters of the SIE plus ES model 
are the angular scale of the critical curve or the 
mass scale inside the critical
curve $b'$, the ellipticity $e$ of the lens and its
position angle $\theta_e$, the strength and the direction of the external shear
$(\gamma,\theta_\gamma)$, the primary lens position
on the lens plane $(x_0,y_0)$, and the image position $(x_i,y_i)$.  
The angles $\theta_e$ and $\theta_\gamma$ are measured East of North
expressed in the observer's coordinates.

It should be emphasized that the observed MIR 
flux ratios are not used for making best-fit lens models. They are used for
only estimating amplitudes of the expected flux anomalies $\eta$. 
To find a set of best-fit parameters,
we use a numerical code GRAVLENS \footnote{See
http://redfive.rutgers.edu/~keeton/gravlens/} developed by Keeton in order to implement
the $\chi^2_{pos}$ fitting of the positions, which have 10
degrees of freedom (8 for quadruple lensed images and 2 for the
center of the primary lens). Because the SIE plus ES (SIE-ES) model has
9 degrees of freedom, residual degree of freedom (dof) is 1.  
If $\chi^2_{pos}/\textrm{dof}<2$ cannot be satisfied, we consider either 
a contribution from
a luminous dwarf galaxy X modeled by an SIS with an Einstein radius
$b_X$ in the neighborhood of
the primary lens (SIE-ES-X) or introduce a large
error for the position of the primary lens in order to satisfy a condition 
$\chi^2/\textrm{dof}<2$ (SIE-ES+).
The latter procedure may be verified in some lens systems 
because any unresolved luminous dwarf galaxies or
inhomogeneous structures inside the galactic bulge of lens galaxy would shift
the position of the center-of-light from the center of the lens
potential. 

In what follows, we briefly review our sample of QSO-galaxy lenses
and the best-fit models. 

\subsection{B1422+231 }
This is a cusp caustic lens that produces 
three colinear bright images A, B, and C with an image opening angle of
$77^\circ .0 $ and a faint image D. The source 
is near a cusp in the astroid-shaped caustic.
The observed MIR flux ratios gives $R_{\textrm{cusp}}=0.20$ 
\citep{chiba2005}. This lens system is the first example that shows a violation 
of the cusp caustic relation \citep{mao1998}. However,
subsequent analysis revealed that the violation is not significant 
when marginalized over the opening angle and the maximum separation between the three
images \citep{keeton2003}.  
The redshift of the source $z_S=3.62$ \citep{kundic1997a} is largest in our 6 samples
and the primary lens is possibly an elliptic galaxy at 
$z_L=0.34$ \citep{tonry1998}. Although the positions of images and lens can be
well fit by the SIE-ES model \citep{chiba2002}, the MIR flux ratios between
the images A, B,
and C are not
consistent with the model prediction \cite{chiba2005}.   
We have confirmed these results and found that inclusion of $m=3$ term
with the external shear does not improve the fit to the MIR flux ratios.
Therefore, it is likely that this lens system is perturbed by 
matter fluctuations on scales smaller than the separation between the images.
Comparing the estimated flux ratios (table 2) to observed ones (table
1), it seems that image A, which is a minimum point in the time arrival surface 
is most likely to have been magnified by perturbers. 
\subsection{  MG0414+0534 }
This is a fold caustic lens where a source is placed near an
astroid-shaped caustic but not near a cusp in the caustic. 
The source at a redshift of $z_S=2.639$ is lensed by a
foreground elliptical galaxy at $z_L=0.9584$ \citep{lawrence1995, tonry1999}. It consists of two close bright images 
A1 and A2 separated by $0''.415$ and two fainter images B and C. 
We have found that the SIE-ES model does not give a good fit to the
data. In order to improve the fit, we have 
considered a possible luminous satellite, object X \citep{schechter1993},
as has been studied in previous lens models (e.g., \citet{Ros2000}). 
The object X is modeled by an SIS at $(x_X, y_X)=(0''.857, 0''.180)$
with an error of $0''.01$ as taken from CASTLES. The SIE-ES-X model
yields a good fit to the positions of images and lens with $\chi^2=0.00
3$. However, the MIR flux ratio of A1 to A2 is not consistent with the
model prediction. The discrepancy remains even
multipoles with $m=3$ and $m=4$ terms are included \citep{minezaki2009}.  
Therefore, it is likely that this lens system is perturbed by 
matter fluctuations on scales smaller than the separation between images A1 and A2. 
Comparing the estimated flux ratios (table 2) to observed ones (table
1), it seems that image A2 which is a saddle point 
in the time arrival surface is most likely to have
been demagnified by perturbers. 
\subsection{  H1413+117    }
This is a ``cross'' lens in which  
quadruply lensed images have an approximate
D4 (dihedral group with 4 rotational symmetries) 
symmetry. It consists of 4 images, A, B, C, and D. 
The source redshift is $z_S=2.55$ \citep{magain1988}
but the lens redshift is unknown. We use the data of the MIR flux ratios
and the lens position observed by \citep{macleod2009}. 
We have found that the SIE-ES model yields a good fit $\chi^2/dof=1.5$ 
to the data of image and lens positions. However, this model
gives a poor fit to the data of the flux ratios. The origin of anomalous
flux ratios may be substructures inside the primary lens or
clustering halos in the line-of-sight. To circumvent this problem,  
\citet{macleod2009} added an SIS at the position of 
another galaxy G2 lying at ($-1.''87, 4''.14$) with
an error of $0.07''$ from image A, 
which corresponds to object \#14 in \citet{kneib1998}.
They have found that $\chi^2/dof\sim 1$ for image and lens 
positions, flux ratios, and weak priors for the lens parameters. 
\citet{goicoechea2010}, found that time delays between images A-D are
also consistent with this model SIE-ES-G2. Assuming a concordance cosmology 
with the Hubble constant $H_0=70 \textrm{km}
\textrm{s}^{-1} \textrm{Mpc}^{-1}$ and density parameters
$\Omega_m=0.3$ and $\Omega_\Lambda=0.7$, they estimated 
the redshift of the primary lens as $z_l=1.88^{+0.09}_{-0.11}$.
Although it is not clear whether $G2$ is the only component that 
would reproduce the observed flux rations, we adopt the SIE-ES-G2
model where the position of the center of an SIS is fixed to 
($-1''.87, 4''.14$) with respect to image A.
\subsection{ PG1115+080   }
This is a fold caustic lens. The 
source at a redshift of $z_S=1.72$ is lensed by a
foreground galaxy at $z_L=0.31$ 
\citep{kundic1997b}. It consists of two close bright images 
A1 and A2 separated by $0''.482$, and two fainter images B and C. 
The data of position of images and lens are taken from CASTLES. 
We find that the SIE-ES model does not provide a good fit to the data
unless the error in lens position is increased from $0.003''$ to
$0.02''$ (SIE-ES+). Interestingly, the value is the same as the
$1\sigma$ error of lens position obtained by 
\citep{macleod2009} for H1413+117. This may be due to systematic
problems in determining the position of the faint lens.
As shown by \cite{chiba2005},
the SIE-ES+ model gives a good fit to the positions of the images and
the lens and the MIR flux ratios \footnote{Alternatively, we may
consider a contribution from a nearby group \citep{sluse2012}.}. 
\subsection{Q2237+0305}
This is the nearest lens in our sample with a 
``cross'' configuration of four images. The source is located at
$z_S=1.695$ and the lens at $z_L=0.0394$ \citep{huchra1985}.
As shown by \citet{minezaki2009}, 
the SIE-ES model gives a good fit to the positions of the images and
the lens as well as the flux ratios. 
\subsection{ RXJ1131-1231}
This is a cusp caustic lens with a source 
at a redshift of $z_S=0.658$ lensed by a
foreground galaxy at $z_L=0.295$ \citep{sluse2003}. 
Unfortunately, no data of fluxes due to 
the MIR continuum emission is available.
Instead, we use the data of fluxes of  
[OIII] emission line from a narrow-line region (NLR)
around the source QSO \citep{sugai2007}.
We find that the SIE-ES model does not provide a good fit to the data
of positions of images and lens unless the error in the position of 
the primary lens is increased from $0.003''$ to $0.017''$ (SIE-ES+). 
We find that the SIE-ES+ model gives a good fit to the positions 
of lensed images 
but the fit to the flux ratios is turned out to be not sufficiently good.
Because the size of the NLR $\sim 100\,\textrm{pc}$ is significantly larger
than the size of the MIR (near IR in rest frame) continuum emission region
($\sim 1 \,\textrm{pc}$), we need a careful consideration on the finite 
source-size effect. In fact, \citet{sugai2007}
found a possible imprint of an extended 
structure in the NLR region. The fractional contribution from the 
extended components can be $\sim 20\%$ for an aperture of $0''.77$
for lensed QSO images. The observed ``bridge'' between
image A and image C and a shift of image B in the opposite direction to
the critical curve suggest an asymmetric structure of the source.  
If this effect is taken into account, then the fit to the flux ratios might
be improved.  This is because the differential magnification of images A and C can lead to 
a reduction in the flux ratio $|\mu_A/\mu_C|$ as extended components of
the images A and C are nearer to the caustic than the corresponding 
core components. Thus we expect additional $\sim 20\%$ uncertainty in the
observed flux ratios.

\section{Flux-ratio anomaly}
\begin{table*}
\hspace{-5.5cm}
\begin{minipage}{136mm}
\caption{Observed MIR Flux Ratios}
\label{symbols}
\begin{tabular}{@{}lccccccccc}
\hline
\hline
Lens & $z_L$ & $z_S$ & $N$ &\multicolumn{3}{c}{Flux Ratio} & $\langle
 \epsilon \rangle  \,('')$
 &  $\langle \theta \rangle  \, ('')$ & Reference  \\
\hline
RXJ1131-1231($\star$) & & & & A/B & C/B &
\\
& 0.295 & 0.658 & 3 &  $1.63^{+0.04}_{-0.02}$  & $1.19^{+0.03}_{-0.12}$
		     & & 0.017 & 1.9 &  1, 2 
\\
\hline
Q2237+0305  & & & & B/A & C/A & D/A &
\\
& 0.04  & 1.695 & 4 & $0.84\pm 0.05$ & $0.46 \pm 0.02$ & $0.87 \pm 0.05$
 & 0.0046 & 0.9 & 1, 3
\\
\hline
PG1115+080 & & & & A2/A1 & & &
\\
& 0.31 &1.72 & 2&  $0.93 \pm 0.06$ &  &  & 0.020 & 1.2 &  1, 4
\\
\hline
H1413+117 & & & & B/A & C/A  & D/A &
\\
& 1.88($\star\star$) &2.55 &4&  $0.84 \pm 0.07$ & $0.72 \pm 0.07 $  & $0.40 \pm 0.06 $
 & 0.020  & 0.6 & 5
\\
\hline
MG0414+0534 & & & & A2/A1 & B/A1  &
\\
 &0.96 &2.639 &3 & $0.90 \pm 0.04$ & $0.36 \pm 0.02 $  &  &0.0042  &1.2
 & 1, 3
\\
\hline 
B1422+231 & & & & A/B & C/B  & &
\\
&0.34 & 3.62 &3 & $0.94 \pm 0.05$ & $0.57 \pm 0.06  $  &  & 0.0042 &1.1
 & 1, 4  
\\ 
\hline
\end{tabular}
References:1. CASTLES; 2. Sugai et al. 2007; 3. Minezaki et al. 2009;
 4. Chiba et al. 2005; 5. MacLeod et al. 2009
\\ Note: ($\star$): [OIII] line flux
 ratios. ($\star \star$): The lens redshift $z_L$ is obtained from a best-fit model using 
the observed positions of the images and the primary lens, the flux ratios, and
the time-delays between the images assuming $H_0=70\,\textrm{km}/\textrm{s}/\textrm{Mpc}$. 

\end{minipage}
\end{table*}
\begin{table*}
\hspace{-5.5cm}
\begin{minipage}{136mm}
\caption{Best-fit Model Parameters and Flux Ratios}
\label{symbols}
\begin{tabular}{@{}lccccccccccccc}
\hline
\hline
Model & $b'$ & ($x_0,y_0$) & $e$ & $\theta_e$ & $\gamma$ & $\theta_\gamma$ 
& $b_\textrm{X}$ & $dof$ & $\chi^2_{pos}$ 
&\multicolumn{3}{c}{Flux Ratio} &$\langle \mu \rangle  $  \\
&($''$) &($''$) & &(deg) & &(deg) & ($''$)&($''$) & & & & \\
\hline
RXJ1131-1231 & & & &  & & & & & & A/B & C/B &
\\
SIE-ES+ & 1.83 & (2.039, 0.568) & 0.145
& -57.8& 0.120 & -81.8 &  & 1 & 1.3 & 1.66 & 0.909 & & 14.6\\
\hline
Q2237+0305 & & & & & & & & & & B/A & C/A & D/A
\\
SIE-ES & 0.854 & (0.075, 0.939) & 0.371
& 64.9 & 0.015 & -46.8 &  & 1 & 0.004& 0.887& 0.447 & 0.825 & 3.73 \\
\hline
PG1115+080 & & & & & & & & & & A2/A1 &  & 
\\
SIE-ES+& 1.14  & (-0.361, -1.342) & 0.156 
& -83.0 & 0.110 & 51.8 & &  1 & 1.0 & 0.912 & & & 12.5 \\
\hline
H1413+117($\star$) & & & & & & & & & & B/A & C/A  & D/A
\\
SIE-ES-X & 0.561 & (-0.172, -0.561) & 0.204 & -14.5 & 0.062
& 55.7 & 0.583 & 2 & 2.2 & 0.894 & 0.905 & 0.458 & 5.24
\\
\hline
MG0414+0534($\star \star$) & & & & & & & & & & A2/A1 & B/A1  & 
\\
SIE-ES-X & 1.08 & (0.472,-1.277)  & 0.232 & -82.1  & 0.102
& 53.8 & 0.185 & 0 & 0.003 & 1.039 & 0.329 & & 13.1 
\\
\hline 
B1422+231 & & & & & & & & & & A/B & C/B  & 
\\ 
SIE-ES &0.755 & (-0.741,-0.658) & 0.309 & -56.6 & 0.166 & -52.3 
& & 1 & 0.55  & 0.797 & 0.512 &  & 4.91 \\
\hline
\end{tabular}
\medskip
\\
($\star$) Object X is modeled by an SIS whose position is fixed at 
 $(x_X,y_X)=(-1.''87,4''.14)$.
\\
($\star \star$) Object X is modeled by an SIS whose position is fitted
to $(x_X,y_X)=(0.''857,0''.180)$ with an error $0''.01$.
\end{minipage}
\end{table*}
\begin{figure}
\vspace*{0cm}
\includegraphics[width=85mm]{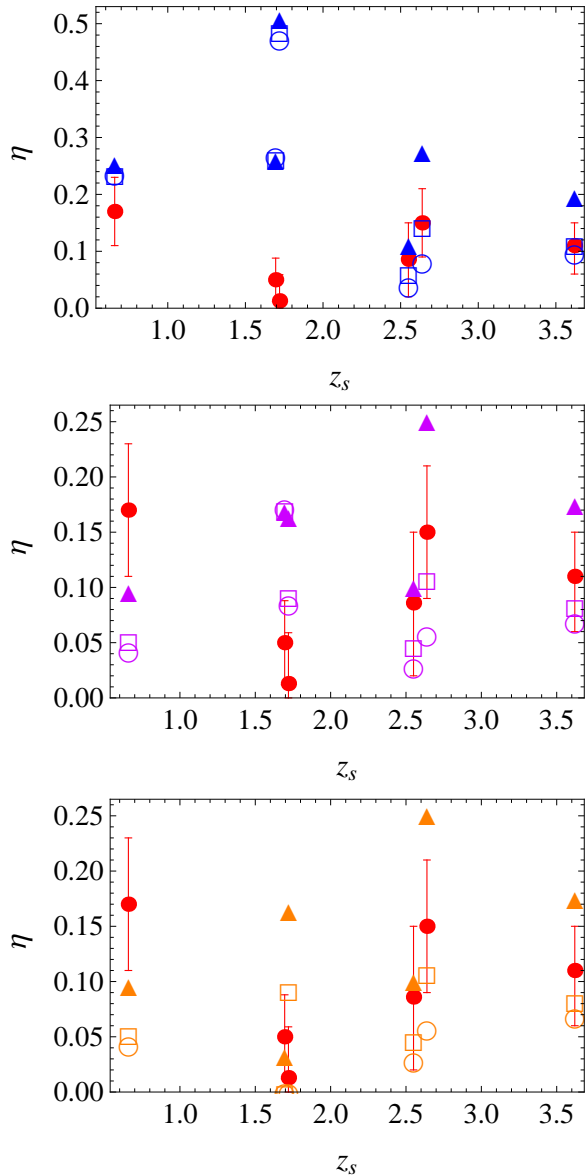}
\caption{Plots of $\eta$ as a function of a source redshift $z_S$ 
for a sample of 6 MIR lenses (disk) and their
predicted values for $k_{max}
=320\, h{\rm Mpc}^{-1}$ (circle),   
$k_{max}=1000 \,h{\rm Mpc}^{-1}$ (square), $k_{max}=10000\,
 h{\rm Mpc}^{-1}$ (solid triangle). Top: we assume $\epsilon=\langle
 \epsilon \rangle$, where $\langle \epsilon \rangle $ is the mean error
in relative position of an image and a lens defined for $N$ images.
 (table 1). Middle: we assume $\epsilon=0.003''$, a typical value for
a lensed image observed in the optical/IR band. Bottom: we assume a cut
 off $k_{lens}$ due to lens modeling and $\epsilon=0.003''$. The error
 bars show the $1\sigma$ errors in the observed MIR fluxes.   }
\label{f12}
\end{figure}
\begin{figure}
\vspace{0.15cm}
\includegraphics[width=85mm]{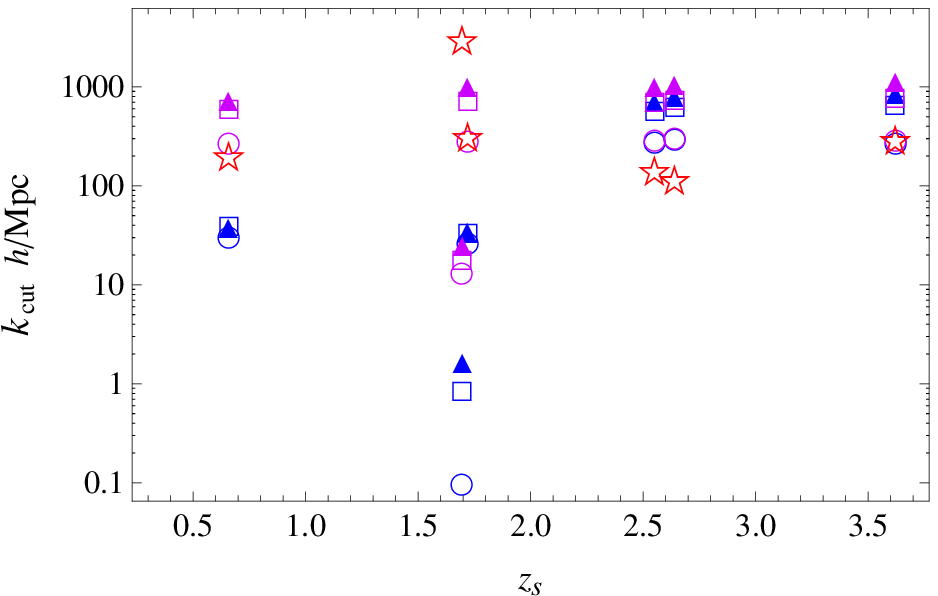}
\caption{Plots of cutoff scales $k_{cut}$ and the source redshift $z_S$ for $\epsilon$ in table  (lower, blue)
and for $\epsilon=0.003''$ (upper, violet) and $k_{lens}$ (star, red).   
The ``UV cut off'' scales are assumed to be
$k_{max}=320 \,h{\rm Mpc}^{-1}$ (circle),   
$k_{max}=1000 \,h{\rm Mpc}^{-1}$ (square), $k_{max}=10000 \,h{\rm Mpc}^{-1}$ (solid triangle).  }
\label{f13}
\vspace{-0.45cm}
\vspace*{1.0cm}
\includegraphics[width=85mm]{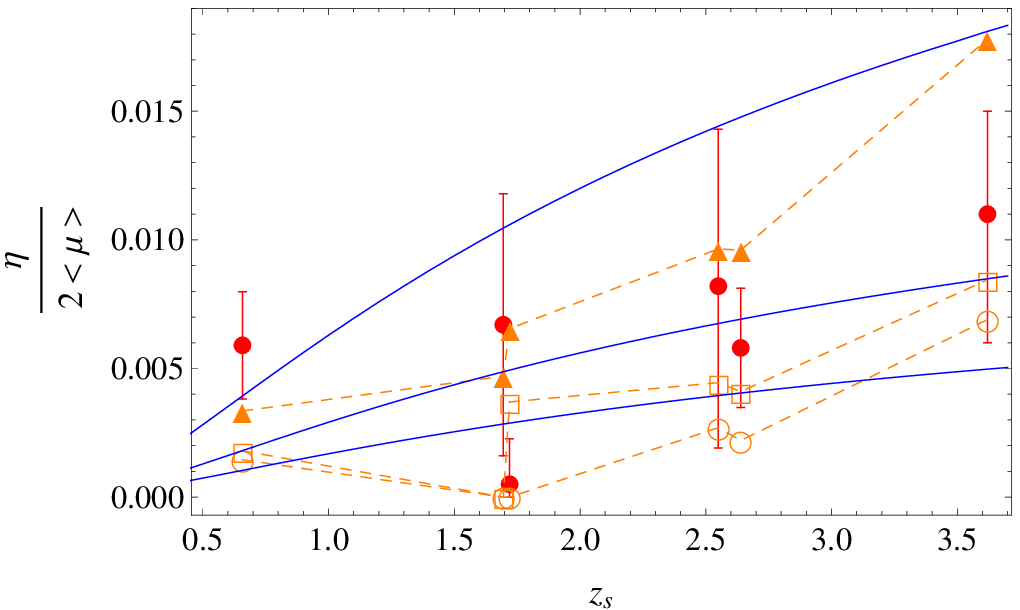}
\caption{Plots of approximated amplitude of convergence
$\sigma_\kappa(0)\sim \eta/2\langle \mu \rangle $ 
as a function of a source redshift $z_S$ 
for the observed MIR lenses (disk) and their
predicted values for $k_{max}=320\, h{\rm Mpc}^{-1}$ (circle),   
$k_{max}=1000\, h{\rm Mpc}^{-1}$ (square), $k_{max}=10000\,
 h{\rm Mpc}^{-1}$ (solid triangle) assuming $k_{cut}$ that corresponds to
 $\epsilon=0.003''$  and cut off $k_{lens}$ due to lens modeling. The error
 bars show the $1\sigma$ errors in the observed MIR fluxes. Assuming
 $z_L=0.5$, rigorous values
 of $\sigma_{\kappa}(0)$ are plotted as full curves for $(k_{cut},k_{max})
=(300\,h{\rm Mpc}^{-1},320\,h{\rm Mpc}^{-1})$ (bottom),
$(750\,h{\rm Mpc}^{-1},1000\,h{\rm Mpc}^{-1})$ (middle), and
 $(1100\,h{\rm Mpc}^{-1},10000\,h{\rm Mpc}^{-1})$ (top).
  }
\label{f14}
\vspace*{0.5cm}
\end{figure}  

 In order to measure a possible contribution of clustering halos in the
 line-of-sight to the flux ratios, we calculate $\eta$ defined in equation (\ref{eta}) 
for our sample of 6 MIR lenses (5 continuum and 1 line emission). 
We assume that the errors of the flux ratios and the 
positions of images and lenses 
obey the Gaussian statistics. We also assume that non-perturbed lens
 potentials are given by best-fit models using 
observed positions of images and lenses. 
We do not consider any contribution from
substructures within a primary lens.  
Effects of image shifts on the flux ratios  
due to intervening halos are assumed to be sufficiently
small though this assumption might not be valid if the allowed shift is
as large as $\epsilon \sim 0.02''$. For simplicity, we use the sharp k-space filter
for determining the maximum scale of fluctuations 
that can affect the flux ratios from errors $\epsilon$ in the relative positions
between an image and the center of the primary lens. To obtain the ``IR'' cutoff $k_{min}$, we use a mean 
separation angle $\langle \theta \rangle $ of an image and the center of the primary lens 
and a mean error
$\langle \epsilon \rangle $ in relative positions of an image and 
the center of the primary lens obtained from quadruple images for each system.

We consider three types of the ``UV'' cutoff, $k_{max}=320, 1000, 10000\, h{\rm Mpc}^{-1}$. 
$k_{max}=320\, h{\rm Mpc}^{-1}$ corresponds to the Nyquist frequency
 $k_{\rm Nyq}$ of our $N$-body simulation. 
For small scale fluctuations 
with wavenumber $k>k_{\rm Nyq}$, we extrapolate the power spectrum obtained
in larger scales $k<k_{\rm Nyq}$. It should be noted that the extrapolated 
power spectrum may be systematically larger/smaller than the correct value for
 $k>k_{\rm Nyq}$. $k_{max}=10000\, h{\rm Mpc}^{-1}$ corresponds to 
an Einstein radius $O[1]\,\rm{pc}$ if the corresponding fluctuation
forms a point mass. As the source sizes of our MIR samples are 
$O[1]\,\rm{pc}$, the contribution of modes $k>10000\, h{\rm Mpc}^{-1}$ is
expected to be negligible.

  As shown in Fig. \ref{f12} (top panel), we find that clustering halos with a mass scale
  of $M\lesssim O[10^7\,\ms]$ or $k>200\, h{\rm Mpc}^{-1}$ in
 the line-of-sight are sufficient in explaining the observed anomalies in the flux
 ratios. However, for Q2237+030 and
 PG1115+080, the predicted anomalies seem too large unless the rms value 
 of $\eta$ is at least comparative to the mean. The apparent discrepancy may be due
 to a possible systematic error in the position of the center of a faint
 lensing galaxy, which is significantly larger than that of lensed
images. A larger error $\epsilon$ tends to give a larger effect on flux
ratios. In order to see this systematic effect, we also calculate
$\eta$ assuming only errors in lensed image positions.
In fact, this assumption is reasnable for estimating the abundance
of possible line-of-sight halos that reside at 
the background of lensing galaxy as the faint lensed image of 
maximum point has not been obeserved in the 6 lenses.
Assuming that $\epsilon= 0.003''$ for all the 6 lenses (``constant $\epsilon$''), 
as shown in Fig. \ref{f12} (middle panel), we find that the fit
to the data of PG1115+080 is
greatly improved. However, the fit to the data of Q2237+030 is not improved.          
This is because the lens redshift is exceptionally small ($z_L=0.04$)
as compared to other 5 lenses in which $z_L \approx 0.3-2.0$. 
As shown in Fig. \ref{f13},
the cut off scale for Q2237+230 is $k_{cut}=O[1-10]\, h{\rm Mpc}^{-1}$. This corresponds to 
mass scales of $10^{11-13}\, \ms$, which are similar to mass scales of
the primary lens.  Therefore, the constraint from the shift
of positions is not so stringent. However, we also need to take into account
the effect of lens modeling as well. Because a constant convergence and a
constant shear of the primary lens are already taken into account in our model,
we need to cut off modes that are equal to or larger than the size of the primary lens.
For simplicity, we assume that modes with a half wavelength 
longer than twice the comoving radius of the critical curve $r(z_L)
\theta_E$ of the primary lens
are cut off, where $r(z_L)$ is the comoving distance to 
the lens at a redshift of $z_L$. The corresponding cut off wavenumber 
is $k_{lens} \equiv 2 \pi/L_{lens}$ where $L_{lens} \sim 4 r(z_L) \theta_E$.   
$\theta_E$ can be estimated as a mean separation angle $\langle \theta
\rangle $ between an
image and the center of the primary for $N$ images.
As we can see in Fig. \ref{f13}, the cut off wavenumbers are $k_{lens}\sim 3100\, h{\rm Mpc}^{-1}$ for 
Q2237+030 and  $k_{lens}=O(10^2)\, h{\rm Mpc}^{-1}$ for 
other 5 systems. This means that the modeling effect 
is significant for systems in which the lens redshift is exceptionally small. 
Taking into account the modeling effect in addition to an assumption 
on the shifts of images and lens $\epsilon=0.003'' $ positions, we find that
the expected $\eta$ for Q2237+030 is significantly
reduced (Fig. \ref{f12}, bottom panel). Furthermore, the predicted values for the other 5 lenses
agree with the data at $\sim 2\, \sigma$ level
without consideration of run-to-run variance of $\eta$.
This result does not change even if we use conservative 
values $\epsilon=\langle \epsilon \rangle$ as shown in table 1. 
Moreover, if we cut off the small scale modes 
$k>k_{lens}$, the effect of differential magnification due to shifts of images
is negligible in comparison with the flux changes due to the weak lensing 
since the magnification perturbation due to the shifts of images is given by
$\delta \mu/\mu\sim \delta r/r_E=O(0.001)$, where
the angular size of the lens is $r_E/r_L\sim 1''$
and the order of the image shift is $\delta r /r_L\sim O(0.001'')$.  

The result suggests that the flux ratio anomalies are caused by 
the weak lensing effect due to extragalactic halos 
with a mass $M \lesssim 10^7\, \ms$ in the line-of-sight. 
In order to see the source redshift $z_S$ dependence of the flux ratio
anomalies, we plot the approximated rms amplitude of convergence due to
intervening halos 
\BE
\sigma_\kappa(0) \sim \frac{\eta}{2 \langle \mu \rangle},
\EE
where $\langle \mu \rangle $ is the mean magnification
obtained from the observed $N$ images for a best-fit model (see table
2). 
As shown in Fig. \ref{f14},
the estimated $\sigma_\kappa(0)$ from the observed MIR flux ratios
monotonically increases as the source redshift $z_S$ increases and it
agrees well with theoretical prediction at $\sim 2\, \sigma$ level.
It should be noted that the expected amplitude of convergence 
$\sigma_\kappa(0)$ for RXJ1131-1231 which shows a deviation
at $\sim 2\, \sigma$ level might decrease if the finite source-size effect is taken
into account as we discussed in section 6 (see also \citet{sugai2007}). 
\section{Conclusions and discussion}

We have studied the weak lensing effect by line-of-sight halos
and sub-halos with a mass of $M \lesssim 10^7\,\ms$ in QSO-galaxy strong
lens systems with quadruple images in a concordant $\Lambda$CDM
universe. Using a polynomially fitted non-linear power spectrum
$P(k)$ obtained from $N$-body simulations that can resolve
halos with a mass of $M \sim 10^5 \ms$, or structures
with a comoving wavenumber $k = 3.2 \times 10^2\, h{\rm Mpc}^{-1}$, we find
that the ratio of magnification perturbation due to intervening
halos to that of a primary lens is typically $\eta  \sim 0.1 $ and
the predicted values agree with the estimated values for 6
QSO-galaxy lens systems (continuum emission for 5 lenses,
line emission from NLR for 1 lens) with quadruple images
in the mid-infrared band without considering the effects of
substructures inside the primary lens. The estimated amplitudes
of convergence perturbation for the 6 lenses increase
with the source redshift as predicted by our semi-analytical
model. This feature strongly supports a hypothesis that the
observed flux ratio anomalies are caused by intervening halos
rather than substructures associated with the primary
lens. However, we do not exclude minor effects from substructures
especially for systems with low lens redshift $z_L$ in
which the weak lensing effect is small. Using an extrapolated
matter power spectrum, we have demonstrated that small
halos with a mass of $M=10^3-10^7 \ms$  can significantly
affect the magnification ratios of lensed images.

Instead of mass $M$, we have used comoving wavenumber $k$ for
parametrizing cut off scale of matter fluctuations due to 
intervening halos. We have considered two types of cut off,
$k_{cut}$ and $k_{lens}$.
$k_{cut}$ is determined from accuracy in positions of lensed images
and the primary lens since intervening halos would induce 
shifts in relative positions of images. 
$k_{lens}$ is given by the (effective) Einstein radius of 
the primary lens. Because large scale fluctuations
are taken into account as a constant 
convergence and a constant shear in lens models, 
fluctuations that are larger than the Einstein radius should be 
neglected. Neglecting the shift of the 
center of a primary lens, we find that $k_{lens}\lesssim
k_{cut}=O[10^2]\, h{\rm Mpc}^{-1}$ for 5 MIR lenses and
$k_{cut}\ll k_{lens}=O[10^3]\, h{\rm Mpc}^{-1}$ for 1 MIR lens
in our sample. 
 
We have not used the cusp-caustic relation $R_{cusp}$ in order to measure the 
strength of flux ratio anomalies since most of our lens systems have
either a complex structure (a luminous satellite) or a broad opening angle 
$\theta > 30^\circ$. Instead, we have devised a new statistic $\eta$, 
to quantify the magnification perturbation. As we need a detailed lens 
model that fits the observed positions of images and lens, it may sounds
less generic than using $R_{cusp}$. In fact, the mass-sheet degeneracy 
yields ambiguity in estimating the magnification
perturbation. Different models with different radial profiles would 
certainly give different predictions. However, this is not a problem.
As we have discussed, perturbations in convergence and shear $\delta \kappa, \delta \gamma$ 
can be measured from extended
sources surrounding the MIR continuum emitting region \citep{inoue2005b}.     
From observed $\eta$, $\delta \kappa $, and $\delta \gamma$, we would be
able to break the mass-sheet degeneracy. This means such an ambiguity
can be removed by estimating shifts of lensed 
images with spatial structures with respect to unperturbed ones.   

Because we have used a new statistic $\eta$ instead of $R_{cusp}$ 
it is difficult to directly compare our result with previous
studies \citep
{metcalf2005a,xu2012} in which the effect of clustering halos is considered to be minor. 
However, as our numerically obtained non-linear power spectrum incorporates
all the effects of clustering halos and that of their substructures, 
our result indicates that 
clustering effect on mass scales of $M\lesssim 10^7 \ms$ is much 
important than considered in previous studies. In fact, we observed
that our
 new statistic $\eta$ is systematically reduced by $20\sim 30$
per cent for $k_{max} \le 1000\,h{\rm Mpc}^{-1}$ and $z_S>2.6$ if no correlation between lensed images is not taken into
account. Moreover, Xu et al. 2012 considered 
only the case $z_S=2.0$ theoretically though 
$z_S$ in our lens systems varies from $0.658$ to $3.62$. We think that
the restriction on the source redshift is one of the weak point in
their analysis as the source redshift dependence is the most important
factor to probe the contribution from the line-
of-sight halos. We have
first shown that observed MIR 
lenses indeed show lens systems with 
high redshift sources tend to exhibit more anomalous flux ratios than
those with low redshift sources. Omitting effects of source redshift
dependence, clustering of halos tend to reduce the signal of anomalous
flux ratios, on the other hand, neglecting constraints
from astrometric shifts or contribution from a constant convergence and shear 
due to line-of-sight halos (yielding upper limit of mass) tends to increase the
signal. Thus, it is difficult to compare our result with the previous
works in literature though the conclusion may look similar.

In order to estimate the magnification perturbation constrained from
shifts of positions of images and lens, we have considered a ``sharp k-space
filter'' for cutting off the fluctuations on large scales. If we use
``Gaussian filters'' that are sufficiently smooth, variance in convergence
can be systematically decreased than using the ``sharp k-space filter''. 
However, if we also consider a 
cut off due to modeling of a primary lens, such an effect may be negligible
as large scale modes are taken into account as a constant convergence or 
shear. It should be noted that we have neglected effects of 
3 point or 4 point correlation of matter fluctuations, which may enhance
the flux ratio anomalies. 
In order to incorporate these effects and check 
validity of our approximation, we need to implement
ray-tracing simulation based on $N$-body simulations. 

If we include effects of baryons, we naively expect further enhancement
in magnification perturbation as baryon cooling would steepen the 
gravitational potential of halos at small scales 
\citep{rudd2008,semboloni2011,vandaalen2011}. 
Then our result would give a lower limit of the amplitude of 
perturbation in magnification ratios. 
However, feedback from supernovae or super massive black holes
could suppress such a steepening near the center of halo due to
outflows \citep{booth2009}. This might eventually
suppress the magnification perturbation due to line-of-sight halos. 
Thus in order to improve our $N$-body simulations
using only collisionless dark matter particles, it 
is very important to incorporate baryonic physics down to 
mass scale of $\sim 10^3 \ms$ or less.
In other words, small scale 
baryonic physics which is relevant to galaxy formation might be
gravitationally probed by the weak 
lensing effect in QSO-galaxy strong lensing system
in the near IR or MIR band.

Next generation telescopes such as the
European Extreme Large Telescope (E-ELT) \citep{gilmozzi2007} 
or Thirty Meter Telescope (TMT) \citep{crampton2006} can
be used to probe hundreds of such strong lens 
systems that are too faint for currently
available largest telescopes to observe. They will 
provide us a unique probe into 
clustering property
of mini-halos with a mass of $M<10^6 \,\ms$.

\section{Acknowledgments}
We acknowledge useful comments from Masashi Chiba, Takeo Minezaki, and
Joseph Silk. We also thank Takahiro Nishimichi for kindly providing us 
the 2LPT code and an anonymous referee for finding various typos in the
manuscript.  This work was supported in part by Hirosaki University Grant for
Exploratory Research by Young Scientists, by the Grant-in-Aid
for Scientific Research on Priority Areas No. 467 ``Probing the
Dark Energy through an Extremely Wide and Deep Survey with
Subaru Telescope'', by the Grand-in-Aid for the Global COE Program
``Quest for Fundamental Principles in the Universe: from Particles
to the Solar System and the Cosmos'' from the Ministry of Education,
Culture, Sports, Science and Technology (MEXT) of Japan, by the
MEXT Grant-in-Aid for Scientific Research on Innovative Areas
 (No. 21111006), by the FIRST program "Subaru Measurements of
Images and Redshifts (SuMIRe)".
Numerical computations were carried out on SR16000 at YITP in Kyoto
University and Cray XT4 at Center for Computational Astrophysics,
CfCA, of National Astronomical Observatory of Japan.

\appendix

\section{Functional Form of Fitting Function}

In this Appendix, we provide the functional form of our fitting function
 of the non-linear matter power spectrum $P(k)$.
The fitting function can be used up to a wavenumber of
 $k = 320\,h{\rm Mpc}^{-1}$, at $0 \leq z \leq 4$ for 
a concordant cosmological model with 
$(\Omega_m, \Omega_b, \Omega_\Lambda,h,n_s,\sigma_8)=
(0.272, 0.046, 0.728, 0.70, 0.97,
 0.81)$, which are obtained from the observed 
WMAP 7yr result \citep{jarosik2011}, the baryon acoustic oscillations (Percival et~al. 2010), 
and $H_0$ \citep{riess2009}.    

The non-linear power spectrum has been frequently calculated using the halo-fit
model by Smith et al. (2003) (hereinafter, S03), which has $30$ fitting
parameters to fit the power spectrum obtained from their $N$-body simulations.
However, as already pointed out by many authors (see, Introduction in
 Takahashi et al. 2012),
 the halo-fit model underestimates the power spectrum in comparison with 
the values from the latest simulations at small scales $k \gtrsim 0.1\,h{\rm Mpc}^{-1}$.
This is because the resolution of simulation in S03 is lower
 than that of the recent simulations.
Recently, Takahashi et al. (2012) have provided an improved halo-fit
model based on the original halo-fit model but re-calculated the fitting parameters
to match their latest simulation results. 
However, their model can be used for only wavenumbers of $k<30\,h{\rm Mpc}^{-1}$, which is
 not sufficient for our purpose. Our interest is in the galactic scale,
 which corresponds to $k > 100\, h{\rm Mpc}^{-1}$.
In this Appendix, we provide a fitting function that can be used to 
calculate the power spectrum at wavenumbers $k\le 320\,h{\rm Mpc}^{-1}$.
Our fitting function is based on the original halo-fit model (S03), but slightly change
 the parameters in their model to fit our simulation results.

In the halo-fit model (S03),
 the dimensionless non-linear power spectrum,
 $\Delta^2(k)=k^3 P(k)/(2 \pi^2)$ consists of one- and two-halo terms: 
\BE
\Delta^2(k) = \Delta_{\rm Q}^2(k) + \Delta_{\rm H}^2(k). 
\EE
The first term is the two-halo term which dominates on large scales,
 while the second term is the one-halo term which dominates on
 small scales. 
We changed the one-halo term to fit our simulation results on small scales.
In S03, there are four parameters $a_n,b_n,c_n,\gamma_n$
 in the one-halo term.
These four parameters are given by as polynomial functions
 of an effective spectrum index $n_{\rm eff}$ and curvature $C$
 calculated by the input linear power spectrum
 (see Appendix of S03 for details).
There are $17$ coefficients in the polynomials,
 and we will determine these $17$ parameters to fit our simulation results.

In order to obtain the fitting function, we use four simulation results:
 three simulations with $L=10\,h{\rm Mpc}^{-1}$ presented in this paper (see section 4),
 and another from the work of Takahashi et al. (2012).
For the four simulations, we adopt the same cosmological model as we
 have mentioned. 
The present three simulations with $L=10\,h{\rm Mpc}^{-1}$ are used for fitting on small scales
 $k>30\,h{\rm Mpc}^{-1}$ while
 the other one is used on large scales $k<30\,h{\rm Mpc}^{-1}$.
The fitting parameters are obtained by using the standard chi-squared analysis.
The chi-square is defined as
\BE
 \chi^2 = \sum_{i} \sum_{k,z} \frac{\left[ P_{\rm model}(k,z) -
 P_{i, {\rm sim}}(k,z)  \right]^2}{2 \sigma_i^2(k,z)},
\label{chi2}
\EE
where $P_{\rm model}$ is the model prediction, and $P_{i, {\rm sim}}$ is
 the four simulation results labeled with an integer $i=1-4$.
We sum up the powers at eight redshifts $z=0,0.35,0.7,1,1.5,2.2,3,4$\footnote{We do
 not use an output at $z=0$ for one simulation of $N_p^3=1024^3$ in
 $L=10h^{-1}$Mpc because of limit in CPU time.} and
 simply set the variance $\sigma_i^2=P_{i, {\rm sim}}^2$ to give an equal
 weight for all the scales.

For the three simulations with $L=10\,h{\rm Mpc}^{-1}$, we sum up $k$ to the Nyquist frequency
 $k_{\rm Nyq}=320(160)\,h{\rm Mpc}^{-1}$ for $N_p^3=1024^3(512^3)$.
On large scales $k \lesssim 10\,h{\rm Mpc}^{-1}$, the sample variance
 among the three simulations is very large (over some tens percent)
 since our simulation box is very small ($10\,h^{-1}$Mpc on a side).
Hence, we use the simulation results in the wavenumber where the
 sample variance is smaller than $16\%$, which corresponds to
 $k>50\,h{\rm Mpc}^{-1}$ for all the redshifts.

We also use the simulation results from Takahashi et al. (2012) on large
 scales $k<30\,h{\rm Mpc}^{-1}$.
Using the same simulation codes as ours, they employed $1024^3$ particles in simulation boxes
 $L=2000,800,320\,h^{-1}$Mpc on a side and combined the $P(k)$ from the
 different box sizes to cover a wide range of scales. 
They prepared $6(3)$ realizations for $L=320(800,2000)\,h^{-1}$Mpc and
  gave the mean power spectrum among the realizations up to
 $k=30\,h{\rm Mpc}^{-1}$.
In order to reduce the shot noise effect, they did not use the 
simulation results at high wavenumbers $k$: the upper limits of the wavenumber are 
 $k_{max}=30\,h{\rm Mpc}^{-1}$ at $z=0, 0.35$, $k_{max}=20\,h{\rm Mpc}^{-1}$ at $z=0.7, 1$,
 $k_{max}=10\,h{\rm Mpc}^{-1}$ at $z=1.5, 2.2$ and $k_{max}=8\,h{\rm
 Mpc}^{-1}$ at $z=3$ in which the power spectrum is $10$ times larger than the shot noise.

Using the standard chi-squared analysis in Eq.(\ref{chi2}), we find
the best-fit parameters:
\\
\BEA
   \log_{10} a_{\rm n} &=& 2.576 +2.263 n_{\rm eff} +1.452 n_{\rm eff}^2
  +0.6308 n_{\rm eff}^3 
\nonumber \\
&+&0.1542 n_{\rm eff}^4 -1.912 C,
\nonumber  \\
   \log_{10} b_{\rm n} &=& 2.062 +1.034 n_{\rm eff} +0.2651 n_{\rm eff}^2
  -3.677 C, 
\nonumber
 \\
\nonumber
   \log_{10} c_{\rm n} &=& 0.4449 +1.743 n_{\rm eff} +0.6772 n_{\rm eff}^2
  +0.06859 C,  \\
   \gamma_{\rm n} &=& 0.2174 -0.1366 n_{\rm eff} +0.2418 C.
\label{bf-params2}
\EEA
The other parameters such as $\alpha_{\rm n},\beta_{\rm n},
 \mu_{\rm n}, \nu_{\rm n},f(\Omega)$ are the same as in S03. 
The definitions of the effective spectrum index $n_{\rm eff}$ and curvature
 $C$ are given in the Appendix of S03.

One can see in Fig.\ref{f2} that our fitting formula agrees well with     
our simulation results. In fact, the root-mean-square 
deviation of our best-fit model of $\Delta^2(k)$ 
from our simulation results is just $5.1 \%$ and the maximum
deviation is $20 \%$ at $k \sim 50\,h{\rm Mpc}^{-1}$ due to the lack of 
available number of modes comparable to the simulation box size.

\bibliographystyle{mn2e}
\bibliography{weak-lensing-by-los}

\begin{thebibliography}{}

\bibitem[\protect\citeauthoryear{Amara, Metcalf, Cox \& Ostriker}{Amara
  et~al.}{2006}]{amara2006}
Amara A.,  Metcalf R.~B.,  Cox T.~J.,    Ostriker J.~P.,  2006, Monthly Notices
  of the Royal Astronomical Society, 367, 1367

\bibitem[\protect\citeauthoryear{{Bartelmann} \& {Schneider}}{{Bartelmann} \&
  {Schneider}}{2001}]{bartelmann2001}
{Bartelmann} M.,  {Schneider} P.,  2001, Physics Reports, 340, 291

\bibitem[\protect\citeauthoryear{{Booth} \& {Schaye}}{{Booth} \&
  {Schaye}}{2009}]{booth2009}
{Booth} C.~M.,  {Schaye} J.,  2009, Monthly Notices of Royal Astronomical
  Society, 398, 53

\bibitem[\protect\citeauthoryear{{Bullock}, {Kravtsov} \& {Weinberg}}{{Bullock}
  et~al.}{2000}]{bullock2000}
{Bullock} J.~S.,  {Kravtsov} A.~V.,    {Weinberg} D.~H.,  2000, Astrophysical
  Journal, 539, 517

\bibitem[\protect\citeauthoryear{{Busha}, {Alvarez}, {Wechsler}, {Abel} \&
  {Strigari}}{{Busha} et~al.}{2010}]{busha2010}
{Busha} M.~T.,  {Alvarez} M.~A.,  {Wechsler} R.~H.,  {Abel} T.,    {Strigari}
  L.~E.,  2010, Astrophysical Journal, 710, 408

\bibitem[\protect\citeauthoryear{Chen}{Chen}{2009}]{chen2009}
Chen J.,  2009, Astronomy \& Astrophysics, 498, 49

\bibitem[\protect\citeauthoryear{Chen, Koushiappas \& Zentner}{Chen
  et~al.}{2011}]{chen2011}
Chen J.,  Koushiappas S.~M.,    Zentner A.~R.,  2011, Astrophysical Journal,
  741, 117

\bibitem[\protect\citeauthoryear{Chen, Kravtsov \& Keeton}{Chen
  et~al.}{2003}]{chen2003}
Chen J.,  Kravtsov A.~V.,    Keeton C.~R.,  2003, Astrophysical Journal, 592,
  24

\bibitem[\protect\citeauthoryear{Chiba}{Chiba}{2002}]{chiba2002}
Chiba M.,  2002, The Astrophysical Journal, 565, 17

\bibitem[\protect\citeauthoryear{Chiba, Minezaki, Kashikawa, Kataza \&
  Inoue}{Chiba et~al.}{2005}]{chiba2005}
Chiba M.,  Minezaki T.,  Kashikawa N.,  Kataza H.,    Inoue K.~T.,  2005,
  Astrophysical Journal, 627, 53

\bibitem[\protect\citeauthoryear{Crampton \& Ellerbroek}{Crampton \&
  Ellerbroek}{2006}]{crampton2006}
Crampton D.,  Ellerbroek B.,  2006, in Whitelock P.,  Dennefeld M.,
  Leibundgut B.,  eds, IAU Symposium NO.232, 2005 Vol.~232, Design and
  development of tmtf.
Cambridge University Press, p.~410

\bibitem[\protect\citeauthoryear{Crocce, Pueblas \& Scoccimarro}{Crocce
  et~al.}{2006}]{crocce2006}
Crocce M.,  Pueblas S.,    Scoccimarro R.,  2006, Monthly Notices of the Royal
  Astronomical Society, 373, 369

\bibitem[\protect\citeauthoryear{{D'Onghia}, {Springel}, {Hernquist} \&
  {Keres}}{{D'Onghia} et~al.}{2010}]{donghia2010}
{D'Onghia} E.,  {Springel} V.,  {Hernquist} L.,    {Keres} D.,  2010,
  Astrophysical Journal, 709, 1138

\bibitem[\protect\citeauthoryear{Eisenstein \& Hu}{Eisenstein \&
  Hu}{1999}]{eisenstein1999}
Eisenstein D.~J.,  Hu W.,  1999, Astrophysical Journal, 511, 5

\bibitem[\protect\citeauthoryear{Gilmozzi \& Spyromilio}{Gilmozzi \&
  Spyromilio}{2007}]{gilmozzi2007}
Gilmozzi R.,  Spyromilio J., , 2007, The 42 m European ELT: status

\bibitem[\protect\citeauthoryear{Goicoechea \& Shalyapin}{Goicoechea \&
  Shalyapin}{2010}]{goicoechea2010}
Goicoechea L.~J.,  Shalyapin V.~N.,  2010, Astrophysical Journal, 708, 995

\bibitem[\protect\citeauthoryear{{Hisano}, {Inoue} \& {Takahashi}}{{Hisano}
  et~al.}{2006}]{hisano2006}
{Hisano} J.,  {Inoue} K.~T.,    {Takahashi} T.,  2006, Physics Letters B, 643,
  141

\bibitem[\protect\citeauthoryear{Huchra, Gorenstein, Kent, Shapiro, Smith,
  Horine \& Perley}{Huchra et~al.}{1985}]{huchra1985}
Huchra J.,  Gorenstein M.,  Kent S.,  Shapiro I.,  Smith G.,  Horine E.,
  Perley R.,  1985, Astronomical Journal, 90, 691


\bibitem[\protect\citeauthoryear{Inoue \& Chiba}{Inoue \&
  Chiba}{2005a}]{inoue2005a}
Inoue K.~T.,  Chiba M.,  2005a, Astrophysical Journal, 633, 23

\bibitem[\protect\citeauthoryear{Inoue \& Chiba}{Inoue \&
  Chiba}{2005b}]{inoue2005b}
Inoue K.~T.,  Chiba M.,  2005b, Astrophysical Journal, 634, 77

\bibitem[\protect\citeauthoryear{Jarosik, Bennett, Dunkley, Gold, Greason,
  Halpern, Hill, Hinshaw, Kogut, Komatsu, Larson, Limon, Meyer, Nolta, Odegard,
  Page, Smith, Spergel, Tucker, Weiland, Wollack \& Wright}{Jarosik
  et~al.}{2011}]{jarosik2011}
Jarosik N.,  Bennett C.~L.,  Dunkley J.,  Gold B.,  Greason M.~R.,  Halpern M.,
   Hill R.~S.,  Hinshaw G.,  Kogut A.,  Komatsu E.,  Larson D.,  Limon M.,
  Meyer S.~S.,  Nolta M.~R.,  Odegard N.,  Page L.,  Smith K.~M.,  Spergel
  D.~N.,  Tucker G.~S.,  Weiland J.~L.,  Wollack E.,    Wright E.~L.,  2011,
  Astrophysical Journal Supplement Series, 192, 1

\bibitem[\protect\citeauthoryear{Keeton, Gaudi \& Petters}{Keeton
  et~al.}{2003}]{keeton2003}
Keeton C.~R.,  Gaudi B.~S.,    Petters A.~O.,  2003, Astrophysical Journal,
  598, 138

\bibitem[\protect\citeauthoryear{Kneib, Alloin, Mellier, Guilloteau, Barvainis
  \& Antonucci}{Kneib et~al.}{1998}]{kneib1998}
Kneib J.~P.,  Alloin D.,  Mellier Y.,  Guilloteau S.,  Barvainis R.,
  Antonucci R.,  1998, Astronomy and Astrophysics, 329, 827

\bibitem[\protect\citeauthoryear{Kormann, Schneider \& Bartelmann}{Kormann
  et~al.}{1994}]{kormann1994}
Kormann R.,  Schneider P.,    Bartelmann M.,  1994, Astronomy and Astrophysics,
  284, 285

\bibitem[\protect\citeauthoryear{Kundic, Cohen, Blandford \& Lubin}{Kundic
  et~al.}{1997}]{kundic1997a}
Kundic T.,  Cohen J.~G.,  Blandford R.~D.,    Lubin L.~M.,  1997, Astronomical
  Journal, 114, 507

\bibitem[\protect\citeauthoryear{Kundic, Hogg, Blandford, Cohen, Lubin \&
  Larkin}{Kundic et~al.}{1997}]{kundic1997b}
Kundic T.,  Hogg D.~W.,  Blandford R.~D.,  Cohen J.~G.,  Lubin L.~M.,    Larkin
  J.~E.,  1997, Astronomical Journal, 114, 2276

\bibitem[\protect\citeauthoryear{Lawrence, Elston, Januzzi \& Turner}{Lawrence
  et~al.}{1995}]{lawrence1995}
Lawrence C.~R.,  Elston R.,  Januzzi B.~T.,    Turner E.~L.,  1995,
  Astronomical Journal, 110, 2570

\bibitem[\protect\citeauthoryear{Maccio \& Miranda}{Maccio \&
  Miranda}{2006}]{maccio2006}
Maccio A.~V.,  Miranda M.,  2006, Monthly Notices of the Royal Astronomical
  Society, 368, 599

\bibitem[\protect\citeauthoryear{McKean, Koopmans, Flack, Fassnacht, Thompson,
  Matthews, Blandford, Readhead \& Soifer}{McKean et~al.}{2007}]{mckean2007}
McKean J.~P.,  Koopmans L. V.~E.,  Flack C.~E.,  Fassnacht C.~D.,  Thompson D.,
   Matthews K.,  Blandford R.~D.,  Readhead A. C.~S.,    Soifer B.~T.,  2007,
  Monthly Notices of the Royal Astronomical Society, 378, 109

\bibitem[\protect\citeauthoryear{MacLeod, Kochanek \& Agol}{MacLeod
  et~al.}{2009}]{macleod2009}
MacLeod C.~L.,  Kochanek C.~S.,    Agol E.,  2009, Astrophysical Journal, 703,
  1177

\bibitem[\protect\citeauthoryear{Magain, Surdej, Swings, Borgeest, Kayser,
  Kuhr, Refsdal \& Remy}{Magain et~al.}{1988}]{magain1988}
Magain P.,  Surdej J.,  Swings J.~P.,  Borgeest U.,  Kayser R.,  Kuhr H.,
  Refsdal S.,    Remy M.,  1988, Nature, 334, 325

\bibitem[\protect\citeauthoryear{Mao \& Schneider}{Mao \&
  Schneider}{1998}]{mao1998}
Mao S.,  Schneider P.,  1998, Monthly Notices of the Royal Astronomical
  Society, 295, 587

\bibitem[\protect\citeauthoryear{Metcalf}{Metcalf}{2005a}]{metcalf2005a}
Metcalf R.~B.,  2005a, The Astrophysical Journal, 629, 673

\bibitem[\protect\citeauthoryear{Metcalf}{Metcalf}{2005b}]{metcalf2005b}
Metcalf R.~B.,  2005b, The Astrophysical Journal, 622, 72

\bibitem[\protect\citeauthoryear{Metcalf \& Amara}{Metcalf \&
  Amara}{2012}]{metcalf2012}
Metcalf R.~B.,  Amara A.,  2012, Monthly Notices of the Royal Astronomical
  Society, 419, 3414

\bibitem[\protect\citeauthoryear{Metcalf \& Madau}{Metcalf \&
  Madau}{2001}]{metcalf2001}
Metcalf R.~B.,  Madau P.,  2001, The Astrophysical Journal, 563, 9

\bibitem[\protect\citeauthoryear{Metcalf, Moustakas, Bunker \& Parry}{Metcalf
  et~al.}{2004}]{metcalf2004}
Metcalf R.~B.,  Moustakas L.~A.,  Bunker A.~J.,    Parry I.~R.,  2004,
  Astrophysical Journal, 607, 43

\bibitem[\protect\citeauthoryear{Minezaki, Chiba, Kashikawa, Inoue \&
  Kataza}{Minezaki et~al.}{2009}]{minezaki2009}
Minezaki T.,  Chiba M.,  Kashikawa N.,  Inoue K.~T.,    Kataza H.,  2009,
  Astrophysical Journal, 697, 610

\bibitem[\protect\citeauthoryear{Miranda \& Maccio}{Miranda \&
  Maccio}{2007}]{miranda2007}
Miranda M.,  Maccio A.~V.,  2007, Monthly Notices of the Royal Astronomical
  Society, 382, 1225

\bibitem[\protect\citeauthoryear{More, McKean, More, Porcas, Koopmans \&
  Garrett}{More et~al.}{2009}]{more2009}
More A.,  McKean J.~P.,  More S.,  Porcas R.~W.,  Koopmans L. V.~E.,    Garrett
  M.~A.,  2009, Monthly Notices of the Royal Astronomical Society, 394, 174

\bibitem[\protect\citeauthoryear{Navarro, Frenk \& White}{Navarro
  et~al.}{1997}]{navarro1997}
Navarro J.~F.,  Frenk C.~S.,    White S. D.~M.,  1997, Astrophysical Journal,
  490, 493

\bibitem[\protect\citeauthoryear{Nishimichi, Shirata, Taruya, Yahata, Saito,
  Suto, Takahashi, Yoshida, Matsubara, Sugiyama, Kayo, Jing \&
  Yoshikawa}{Nishimichi et~al.}{2009}]{nishimichi2009}
Nishimichi T.,  Shirata A.,  Taruya A.,  Yahata K.,  Saito S.,  Suto Y.,
  Takahashi R.,  Yoshida N.,  Matsubara T.,  Sugiyama N.,  Kayo I.,  Jing
  Y.~P.,    Yoshikawa K.,  2009, Publications of the Astronomical Society of
  Japan, 61, 321

\bibitem[\protect\citeauthoryear{{Press} \& {Schechter}}{{Press} \&
  {Schechter}}{1974}]{press1974}
{Press} W.~H.,  {Schechter} P.,  1974, Astrophysical Journal, 187, 425

\bibitem[\protect\citeauthoryear{Riess, Macri, Casertano, Sosey, Lampeitl,
  Ferguson, Filippenko, Jha, Li, Chornock \& Sarkar}{Riess
  et~al.}{2009}]{riess2009}
Riess A.~G.,  Macri L.,  Casertano S.,  Sosey M.,  Lampeitl H.,  Ferguson
  H.~C.,  Filippenko A.~V.,  Jha S.~W.,  Li W.~D.,  Chornock R.,    Sarkar D.,
  2009, Astrophysical Journal, 699, 539

\bibitem[\protect\citeauthoryear{Ros, Guirado, Marcaide, Perez-Torres, Falco,
  Munoz, Alberdi \& Lara}{Ros et~al.}{2000}]{Ros2000}
Ros E.,  Guirado J.~C.,  Marcaide J.~M.,  Perez-Torres M.~A.,  Falco E.~E.,
  Munoz J.~A.,  Alberdi A.,    Lara L.,  2000, Astronomy and Astrophysics, 362,
  845

\bibitem[\protect\citeauthoryear{{Rudd}, {Zentner} \& {Kravtsov}}{{Rudd}
  et~al.}{2008}]{rudd2008}
{Rudd} D.~H.,  {Zentner} A.~R.,    {Kravtsov} A.~V.,  2008, Astronomical
  Journal, 672, 19

\bibitem[\protect\citeauthoryear{Schechter \& Moore}{Schechter \&
  Moore}{1993}]{schechter1993}
Schechter P.~L.,  Moore C.~B.,  1993, Astronomical Journal, 105, 1

\bibitem[\protect\citeauthoryear{{Semboloni}, {Hoekstra}, {Schaye}, {van
  Daalen} \& {McCarthy}}{{Semboloni} et~al.}{2011}]{semboloni2011}
{Semboloni} E.,  {Hoekstra} H.,  {Schaye} J.,  {van Daalen} M.~P.,
  {McCarthy} I.~G.,  2011, Monthly Notices of Royal Astronomical Society, 417,
  2020

\bibitem[\protect\citeauthoryear{{Sheth} \& {Tormen}}{{Sheth} \&
  {Tormen}}{2002}]{sheth2002}
{Sheth} R.~K.,  {Tormen} G.,  2002, Monthly Notices of the Royal Astronomical
  Society, 329, 61

\bibitem[\protect\citeauthoryear{Shin \& Evans}{Shin \& Evans}{2008}]{shin2008}
Shin E.~M.,  Evans N.~W.,  2008, Monthly Notices of the Royal Astronomical
  Society, 390, 505

\bibitem[\protect\citeauthoryear{{Sluse}, {Chantry}, {Magain}, {Courbin} \&
  {Meylan}}{{Sluse} et~al.}{2012}]{sluse2012}
{Sluse} D.,  {Chantry} V.,  {Magain} P.,  {Courbin} F.,    {Meylan} G.,  2012,
  Astronomy and Astrophysics, 538, A99

\bibitem[\protect\citeauthoryear{Sluse, Surdej, Claeskens, Hutsemekers, Jean,
  Courbin, Nakos, Billeres \& Khmil}{Sluse et~al.}{2003}]{sluse2003}
Sluse D.,  Surdej J.,  Claeskens J.~F.,  Hutsemekers D.,  Jean C.,  Courbin F.,
   Nakos T.,  Billeres M.,    Khmil S.~V.,  2003, Astronomy and Astrophysics,
  406, L43

\bibitem[\protect\citeauthoryear{Smith, Peacock, Jenkins, White, Frenk, Pearce,
  Thomas, Efstathiou \& Couchman}{Smith et~al.}{2003}]{smith2003}
Smith R.~E.,  Peacock J.~A.,  Jenkins A.,  White S. D.~M.,  Frenk C.~S.,
  Pearce F.~R.,  Thomas P.~A.,  Efstathiou G.,    Couchman H. M.~P.,  2003,
  Monthly Notices of the Royal Astronomical Society, 341, 1311

\bibitem[\protect\citeauthoryear{Springel}{Springel}{2005}]{springel2005}
Springel V.,  2005, Monthly Notices of the Royal Astronomical Society, 364,
  1105

\bibitem[\protect\citeauthoryear{Springel, Yoshida \& White}{Springel
  et~al.}{2001}]{springel2001}
Springel V.,  Yoshida N.,    White S. D.~M.,  2001, New Astronomy, 6, 79

\bibitem[\protect\citeauthoryear{Sugai, Kawai, Shimono, Hattori, Kosugi,
  Kashikawa, Inoue \& Chiba}{Sugai et~al.}{2007}]{sugai2007}
Sugai H.,  Kawai A.,  Shimono A.,  Hattori T.,  Kosugi G.,  Kashikawa N.,
  Inoue K.~T.,    Chiba M.,  2007, Astrophysical Journal, 660, 1016

\bibitem[\protect\citeauthoryear{Takahashi, Sato, Nishimichi, Taruya \&
  Oguri}{Takahashi et~al.}{2012}]{takahashi2012}
Takahashi R.,  Sato M.,  Nishimichi T.,  Taruya A.,    Oguri M., 
Astrophysical Journal, submitted (arXiv:1208.2701), 2012

\bibitem[\protect\citeauthoryear{Tonry}{Tonry}{1998}]{tonry1998}
Tonry J.~L.,  1998, Astronomical Journal, 115, 1

\bibitem[\protect\citeauthoryear{Tonry \& Kochanek}{Tonry \&
  Kochanek}{1999}]{tonry1999}
Tonry J.~L.,  Kochanek C.~S.,  1999, Astronomical Journal, 117, 2034

\bibitem[\protect\citeauthoryear{Valageas \& Nishimichi}{Valageas \&
  Nishimichi}{2011}]{valageas2011}
Valageas P.,  Nishimichi T.,  2011, Astronomy and Astrophysics, 527, 87

\bibitem[\protect\citeauthoryear{{van Daalen}, {Schaye}, {Booth} \& {Dalla
  Vecchia}}{{van Daalen} et~al.}{2011}]{vandaalen2011}
{van Daalen} M.~P.,  {Schaye} J.,  {Booth} C.~M.,    {Dalla Vecchia} C.,  2011,
  Monthly Notices of Royal Astronomical Society, 415, 3649

\bibitem[\protect\citeauthoryear{{Vegetti}, {Lagattuta}, {McKean}, {Auger},
  {Fassnacht} \& {Koopmans}}{{Vegetti} et~al.}{2012}]{vegetti2012}
{Vegetti} S.,  {Lagattuta} D.~J.,  {McKean} J.~P.,  {Auger} M.~W.,  {Fassnacht}
  C.~D.,    {Koopmans} L.~V.~E.,  2012, Nature, 481, 341

\bibitem[\protect\citeauthoryear{Wong, Keeton, Williams, Momcheva \&
  Zabludoff}{Wong et~al.}{2011}]{wong2011}
Wong K.~C.,  Keeton C.~R.,  Williams K.~A.,  Momcheva I.~G.,    Zabludoff
  A.~I.,  2011, Astrophysical Journal, 726, 84

\bibitem[\protect\citeauthoryear{Xu, Mao, Wang, Springel, Gao, White, Frenk,
  Jenkins, Li \& Navarro}{Xu et~al.}{2009}]{xu2009}
Xu D.,  Mao S.,  Wang J.,  Springel V.,  Gao L.,  White S.,  Frenk C.,  Jenkins
  A.,  Li G.,    Navarro J.,  2009, Monthly Notices of the Royal Astronomical
  Society, 398, 1235

\bibitem[\protect\citeauthoryear{Xu, Mao, Cooper, Gao, Frenk, Angulo \&
  Helly}{Xu et~al.}{2012}]{xu2012}
Xu D.~D.,  Mao S.,  Cooper A.~P.,  Gao L.,  Frenk C.~S.,  Angulo R.~E.,
  Helly J.,  2012, Monthly Notices of the Royal Astronomical Society, 421, 2553

\bibitem[\protect\citeauthoryear{Xu, Mao, Cooper, Wang, Gao, Frenk \&
  Springel}{Xu et~al.}{2010}]{xu2010}
Xu D.~D.,  Mao S.~D.,  Cooper A.~P.,  Wang J.,  Gao L.~A.,  Frenk C.~S.,
  Springel V.,  2010, Monthly Notices of the Royal Astronomical Society, 408,
  1721

\end{thebibliography}

\end{document}